\documentclass[jgrga]{agutex}
\usepackage{amsmath} 
\usepackage{amssymb} 
\usepackage[]{graphicx}
\authorrunninghead{Komar et al.}
\titlerunninghead{Tests of Dayside Reconnection Models}
\usepackage{verbatim}
\usepackage{mathrsfs}

\begin{document}
\hskip 8.5cm {\it Accepted - December 15, 2014} \vskip .125cm 

\title{Comparative analysis of dayside magnetic reconnection models in
  global magnetosphere simulations}

\authors{C. M. Komar,\altaffilmark{1} R. L. Fermo,\altaffilmark{1}
  P. A. Cassak\altaffilmark{1}}
\altaffiltext{1}{Department of Physics and Astronomy, West Virginia
  University, Morgantown, West Virginia, USA}
\authoraddr{C. M. Komar, Department of Physics and Astronomy, White
  Hall, Box 6315, West Virginia University, Morgantown, WV
  26506. (ckomar@mix.wvu.edu)}

\begin{abstract}

  We test and compare a number of existing models predicting the
  location of magnetic reconnection at Earth's dayside magnetopause
  for various solar wind conditions.  We employ robust image
  processing techniques to determine the locations where each model
  predicts reconnection to occur.  The predictions are then compared
  to the magnetic separators, the magnetic field lines separating
  different magnetic topologies.  The predictions are tested in
  distinct high-resolution simulations with interplanetary magnetic
  field (IMF) clock angles ranging from 30 to 165 degrees in global
  magnetohydrodynamic simulations using the three-dimensional
  Block-Adaptive Tree Solarwind Roe-type Upwind Scheme (BATS-R-US)
  code with a uniform resistivity, although the described techniques
  can be generally applied to any self-consistent magnetosphere code.
  Additional simulations are carried out to test location model
  dependence on IMF strength and dipole tilt. We find that most of the
  models match large portions of the magnetic separators when the IMF
  has a southward component, with the models saying reconnection
  occurs where the local reconnection rate and reconnection outflow
  speed are maximized performing best.  When the IMF has a northward
  component, none of the models tested faithfully map the entire
  magnetic separator, but the maximum magnetic shear model is the best
  at mapping the separator in the cusp region where reconnection has
  been observed.  Predictions for some models with northward IMF
  orientations improve after accounting for plasma flow shear parallel
  to the reconnecting components of the magnetic fields.  Implications
  for observations are discussed.

\end{abstract}

\begin{article}
\section{Introduction}
\label{section::Introduction}
Magnetic reconnection occurs in plasmas where oppositely directed
magnetic fields merge, effectively break and transfer their magnetic
energy into kinetic energy and heat.  This process occurs between the
interplanetary magnetic field (IMF) and the terrestrial magnetic field
at the dayside magnetopause, transferring solar wind plasma into
near-Earth space; this is the first step in the Dungey cycle of
magnetospheric convection~\citep{dungey1961}.  However, predicting
where magnetic reconnection occurs at Earth's dayside magnetopause for
arbitrary solar wind conditions has been the subject of studies for
nearly fifty years and remains a challenging problem.  Knowledge of
where reconnection occurs at Earth's magnetopause is crucial for
developing a quantitative prediction of geomagnetic activity for space
weather applications and to support satellite missions studying
magnetic reconnection, such as NASA's upcoming Magnetospheric
Multiscale (MMS) mission~\citep{burch2009,moore2013}.

In the classical sketches of~\citet{dungey1961,dungey1963},
reconnection occurs at the subsolar point for due southward IMF and
near the polar cusps for due northward IMF orientations (ignoring
Earth's dipole tilt).  However, much less is known about where
reconnection occurs when the IMF makes an arbitrary clock angle
$\theta_\text{IMF}$ with the Earth's dipole axis.  The location of
dayside magnetic reconnection was originally discussed in terms of
anti-parallel~\citep{crooker1979} or component reconnection, either
with a uniform out-of-plane (guide) magnetic
field~\citep{sonnerup1974,gonzalez1974} or equal and opposite
components of the reconnecting magnetic field~\citep{cowley1976}); the
latter is known to not be valid.  A number of literature reviews have
been written on this
subject~\citep{dorelli2007,paschmann2008,fuselier2011,guo2013}; the
upshot is that both component and anti-parallel reconnection have been
seen in observations and simulations.  This implies that neither model
can explain all the data so models going beyond them are necessary.

Some recent models suggest that reconnection occurs where some
parameter is maximized which produces reconnection at an optimum
efficiency.  One such model is the maximum magnetic shear
model~\citep{trattner2007}, an extension of the anti-parallel
reconnection hypothesis.  This model suggests that reconnection occurs
where the magnetic shear angle between the magnetospheric and
magnetosheath magnetic fields is a maximum.  Three related models
suggest reconnection occurs where the reconnection outflow speed is
maximized~\citep{swisdak2007}, the reconnection rate is
maximized~(Shay, private communication, 2009;~\citep{borovsky2013}),
and the reconnecting field's magnetic energy is
maximized~\citep{hesse2013}.  Other models suggest that reconnection
occurs where the current density is a maximum~\citep{alexeev1998} or
where the magnetospheric magnetic field and the draped IMF are
bisected~\citep{moore2002}.

There are a number of studies testing some of these models in
different contexts.  The maximum magnetic shear model was shown to be
consistent with observations of magnetic reconnection at planetary
magnetopauses with data from Cluster~\citep{trattner2007,dunlop2011},
THEMIS~\citep{trattner2012a}, and Cassini~\citep{fuselier2014}.  Also,
observations~\citep{teh2008} and simulations~\citep{schreier2010}
suggest that the maximization of the asymmetric reconnection outflow
speed controls the location and orientation of reconnection, although
it is difficult to distinguish maximum outflow speed from maximum
reconnection rate. The angle of bisection hypothesis has been tested
with observations and in global magnetospheric
simulations~\citep{pulkkinen2010} and in two-dimensional (2D)
reconnection simulations~\citep{hesse2013}.

One reason the location of reconnection at the dayside magnetopause is
difficult to predict is that reconnection does not occur at a single
point, but rather occurs along a line of finite extent.  Magnetopause
reconnection occurs along the line that separates magnetic fields of
different topologies - solar wind and closed magnetospheric field
lines merge and reconnect, resulting in open field lines.  This line
is called the magnetic
separator~\citep{cowley1973,stern1973,siscoe1987,lau1990,siscoe2001,parnell2010},
and is the three-dimensional (3D) analog of the 2D X-line.  There is
observational~\citep{xiao2007,pu2013,guo2013} and
numerical~\citep{hu2004,laitinen2006,laitinen2007,dorelli2007,dorelli2008,dorelli2009,hu2009,pulkkinen2010,ouellette2010,peng2010,cnossen2012,komar2013}
evidence for the existence of separators throughout Earth's
magnetosphere.

Consequently, a careful assessment of the correctness of the recent
models for the location and orientation of reconnection begins with a
comparison of their predictions to the measured magnetic separators.
To do so, we interpret the models, as~\citet{trattner2007} has done
with the maximum magnetic shear model, as predicting the collections
of points that locally maximize a particular quantity rather than the
single point that globally maximizes the quantity in question.

The goal of this study is to systematically test the recent models
against the separators arising in self-consistent global simulations
of the magnetosphere.  To do so, we employ image processing techniques
to unambiguously determine each model's prediction in our
simulations. Each model's prediction is then compared with the
magnetic separator determined with the algorithm described
in~\citet{komar2013}; this algorithm has been shown to work for IMF
with an arbitrary orientation.  The present study tests each model's
agreement with the separators for various IMF clock angle
$\theta_\text{IMF}$, IMF strength $B_\text{IMF}$, and dipole tilt
angle.  We also present a test of one possible way to incorporate the
effects of a flow shear parallel to the reconnecting fields into the
models based on reconnection outflow speed and reconnection rate.

The layout of this paper is as follows:
Section~\ref{section::Reconnection_Location_Models} reviews the models
of dayside reconnection location that we test in the present study.
Section~\ref{section::Methodology} describes our methodology: the
method used to determine the magnetopause, a robust image processing
algorithm to determine each model's prediction, and details of our
global magnetospheric simulations.  Our results are presented in
Section~\ref{section::Results}.  A summary of our results and
applications are discussed in Section~\ref{section::Conclusions}.

\section{Reconnection Location Models}
\label{section::Reconnection_Location_Models}
\begin{figure}[t]
\centering
\noindent\includegraphics[width=20pc]{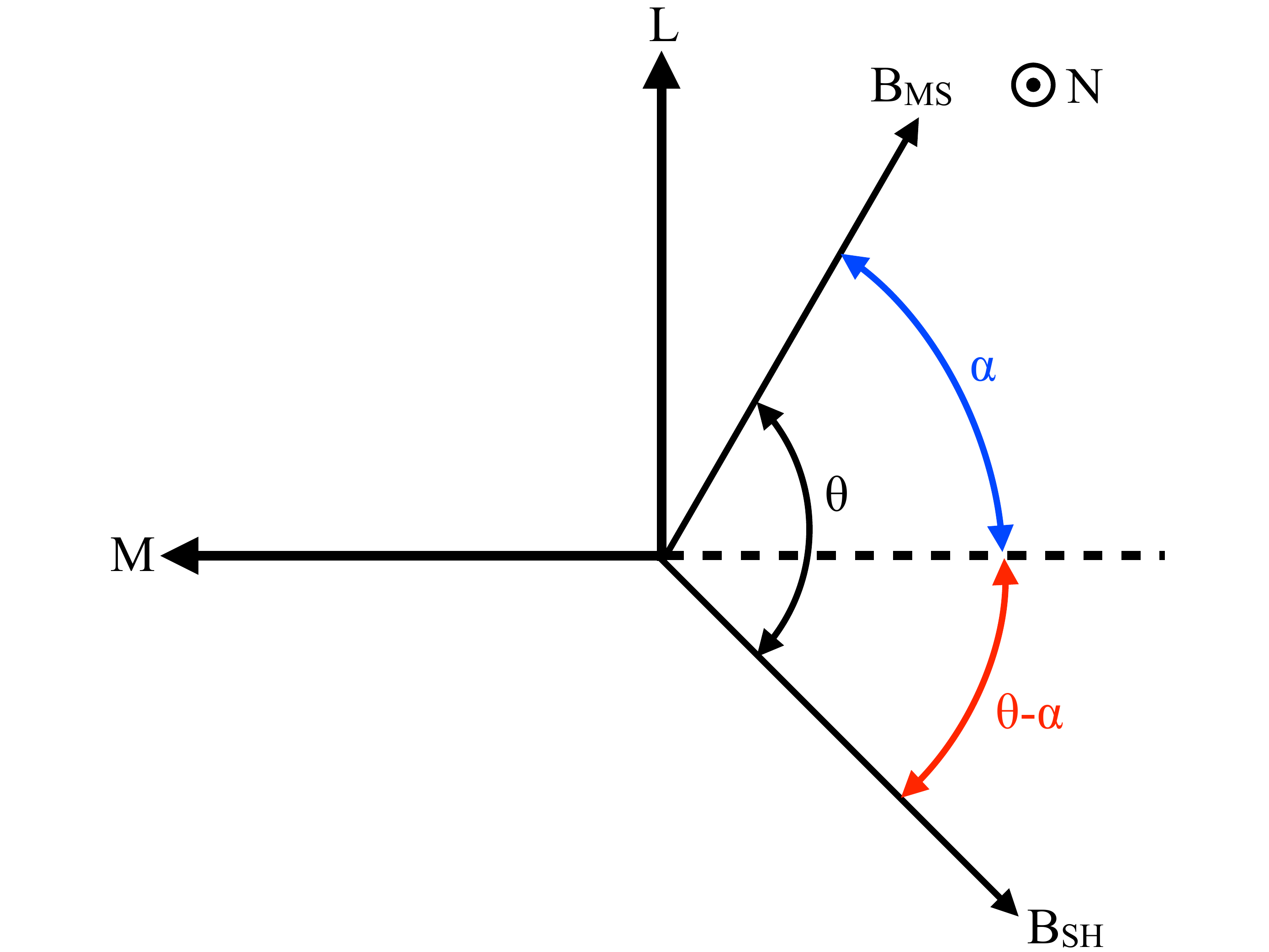}
\caption{Schematic of an arbitrary magnetospheric magnetic field
  $\mathbf{B}_\text{MS}$ making an angle $\theta$ with the
  magnetosheath magnetic field $\mathbf{B}_\text{SH}$ in the boundary
  normal (LMN) coordinate system.  The angle $\alpha$ is the angle the
  magnetospheric field makes with the $M$ axis. Adapted
  from~\citet{sonnerup1974} and~\citet{swisdak2007}.}
\label{fig::LMN_Coordinates}
\end{figure}
This section reviews the models tested in this study.  We employ the
boundary normal (LMN) coordinate system.  The $N$ direction is the
magnetopause normal corresponding to the inflow direction, the $L$
direction corresponds to the reconnecting component of the magnetic
field, and the $M$ direction gives the direction of the out-of-plane
(guide) magnetic field, completing the right-handed triplet.
Figure~\ref{fig::LMN_Coordinates}, adapted from~\citet{sonnerup1974}
and~\citet{swisdak2007}, displays the magnetospheric and magnetosheath
magnetic fields $\mathbf{B}_\text{MS}$ and $\mathbf{B}_\text{SH}$ at
the magnetopause viewed along the inflow direction.  These two fields
point in arbitrary directions in the $L$-$M$ plane, making a magnetic
shear angle of $\theta$ with each other.  The angle $\alpha$ is the
angle $\mathbf{B}_\text{MS}$ makes with the $M$ axis and is
constrained to occur within the range $0<\alpha<\theta$.  It is
tacitly assumed that the magnetic field components along the $N$ axis
are weak.

In the maximum magnetic shear model~\citep{trattner2007}, reconnection
occurs at magnetopause locations where the magnetic shear angle
$\theta$ between the magnetospheric and magnetosheath magnetic fields
is a maximum.  This model identifies the anti-parallel reconnection
locations of~\citet{crooker1979} with high fidelity, while also
identifying additional locations away from the anti-parallel regions
where $\theta$ is locally a maximum.  The magnetic shear angle
$\theta$ is calculated from the magnetospheric and magnetosheath
magnetic fields as
\begin{equation}
  \cos{\theta}=\frac{\mathbf{B}_\text{MS}\cdot\mathbf{B}_\text{SH}}{B_\text{MS}B_\text{SH}}.
\label{eqn::Shear_Angle}
\end{equation}

Alternative explanations have suggested that reconnection occurs at
locations where some measure of reconnection efficiency is maximized.
In the model of~\citet{swisdak2007}, reconnection on the dayside is
oriented (has an angle $\alpha$) to maximize the reconnection outflow
speed $c_{A,out}$. In a subsequent model, reconnection orients so as
to maximize the reconnection electric field $E$~(Shay, private
communication, 2009),~\citep{borovsky2013}).  To calculate the
predictions from these models at the dayside magnetopause, one needs
the plasma parameters of the magnetosheath and magnetosphere, where
asymmetries in the magnetic field and plasma density strongly affect
reconnection.  The scaling of asymmetric reconnection was developed
for 2D antiparallel reconnection using conservation of mass and
energy~\citep{cassak2007c} to obtain an expression for the asymmetric
outflow speed
\begin{equation}
  c_{A, out}^2\sim\frac{B_\text{MS,L}B_\text{SH,L}\left(B_\text{MS,L}+B_\text{SH,L}\right)}{\mu_0\left(\rho_\text{MS}B_\text{SH,L}+\rho_\text{SH}B_\text{MS,L}\right)}
\label{eqn::Asymmetric_Outflow_Speed}
\end{equation}
and for the asymmetric electric field
\begin{equation}
  E\sim c_{A,out}\frac{B_\text{MS,L}B_\text{SH,L}}{B_\text{MS,L}+B_\text{SH,L}}\frac{2\delta}{L},
\label{eqn::Generalized_Asymmetric_Reconnection_Rate}
\end{equation}
where $\rho_\text{MS}$ and $\rho_\text{SH}$ are the upstream
magnetospheric and magnetosheath plasma mass densities, respectively,
and $\delta$ and $L$ are the half-width and half-length of the
diffusion region.
Equation~\eqref{eqn::Generalized_Asymmetric_Reconnection_Rate} is
valid independent of dissipation
mechanism~\citep{cassak2008b,cassak2009c}; the present study employs a
resistivity $\eta$ and the reconnection rate for asymmetric
collisional, or Sweet-Parker, reconnection is~\citep{cassak2007c}
\begin{equation}
  E_{SP}\sim\sqrt{\frac{\eta c_{A, out}}{\mu_0 L}B_\text{MS,L}B_\text{SH,L}}.
\label{eqn::SP_Asymmetric_Reconnection_Rate}
\end{equation}

From Fig.~\ref{fig::LMN_Coordinates}, the reconnecting component of
the upstream magnetospheric magnetic field can be written as
\begin{equation}
  B_\text{MS,L}=B_\text{MS}\sin{\alpha}
\label{eqn::Magnetosphere_L}
\end{equation}
and for the magnetosheath as
\begin{equation}
  B_\text{SH,L}=B_\text{SH}\sin{\left(\theta-\alpha\right)}.
\label{eqn::Magnetosheath_L}
\end{equation}
The asymmetric outflow speed is maximized by setting
\begin{equation*}
  \frac{\partial c^2_{A,\,out}}{\partial\alpha}=0
\end{equation*}
and solving for $\alpha$~\citep{swisdak2007}.  A similar approach is
used to maximize $E$~(Shay, private communication,
2009;~\citep{borovsky2013}).  In practice, we interpret these as
saying that the local $c_{A,out}$ and $E$ can be found at every
location on the magnetopause by finding $\alpha$ at every location,
and that reconnection occurs where the parameter is a local maximum.
We note in passing that this approach assumes that any nonzero $B_M$
component of either magnetic field does not affect the outflow speed
or reconnection rate.  This is not likely to be correct in the real
magnetosphere where finite Larmor radius effects are important, but
should be accurate for the MHD simulations in the present study.
Another limitation of the present formulation of these arguments is
that they neglect the importance of the plasma bulk flow parallel to
the reconnecting magnetic field components; we detail an attempt to
account for this effect and measure the impact on the predictions of
the asymmetric reconnection models in
Section~\ref{subsection::Shear_Flow}.

More recently,~\citet{hesse2013} suggested that the reconnection rate
is proportional to the product of the reconnecting component's
magnetic energies,
\begin{equation}
  E\propto B_\text{MS,L}^2B_\text{SH,L}^2,
\label{eqn::E_Hesse}
\end{equation}
and reconnection maximizes this product.  Applying a maximization
argument with respect to $\alpha$ gives a solution of
$\alpha=\theta/2$; this is equivalent to previous arguments where the
guide field bisects the merging magnetosheath and magnetospheric
magnetic fields~\citep{moore2002,borovsky2008b}.

We also test the model that predicts reconnection to occur where the
current density magnitude $\left|\mathbf{J}\right|$ is a
maximum~\citep{alexeev1998}.  A related approach has been to look at
the divergence of the Poynting vector.  Reconnection has been located
by measuring this quantity in observations~\citep{anekallu2013} and
simulations~\citep{papadopoulos1999,palmroth2003,laitinen2006,laitinen2007,pulkkinen2010,palmroth2012,hoilijoki2014}.
In steady-state, the divergence of the Poynting vector is
\begin{equation*}
  \nabla\cdot\left(\frac{\mathbf{E}\times\mathbf{B}}{\mu_0}\right)\sim-\mathbf{E}\cdot\mathbf{J},
\end{equation*}
where $\mathbf{E}$ is the electric field given by Ohm's law.  The
present study employs a large explicit resistivity, to be described in
Section~\ref{subsection::Simulation_Study}. In this system, the divergence of
the Poynting vector simplifies to $-\eta J^2$ at the reconnection site.  We
therefore argue that, up to a scaling factor, the Poynting vector's
divergence would give similar results as the maximum current density
magnitude argument for this study.

For completeness, we also test the component reconnection hypothesis
as originally proposed by~\citet{sonnerup1974}
and~\citet{gonzalez1974}, even though
observations~\citep{gosling2007a,gosling2007b,teh2008,phan2009,phan2010,gosling2013,dibraccio2013}
and simulations~\citep{schreier2010} suggest the model is not valid.
In this model, the local reconnection geometry orients itself such
that the $M$ component of the magnetic field is uniform, which, using
the notation in Fig.~\ref{fig::LMN_Coordinates}, implies
\begin{equation}
\tan\alpha=\frac{b-\cos\theta}{\sin\theta},
\label{eqn::Sonnerup_Criterion}
\end{equation}
where $b=B_\text{MS}/B_\text{SH}$.  This model places an additional
constraint on the orientation angle $\alpha$, with valid solutions
only when the magnetic shear angle $\theta$ is greater than the
critical shear angle $\theta_c=\cos^{-1}\left(1/b\right)$.

\section{Methodology}
\label{section::Methodology}
Each of the models discussed in the previous section suggest
reconnection at the dayside magnetopause occurs where some parameter
is a local maximum.  We model our approach after~\citet{trattner2007},
who identified a curve corresponding to where the magnetic shear angle
$\theta$ was a local maximum at the magnetopause.  The magnetospheric
magnetic field was modeled using the T96 model~\citep{tsyganenko1995}
and the Cooling model is used for the draped magnetosheath magnetic
field~\citep{cooling2001} to calculate $\theta$ at many locations on
the dayside magnetopause.  The local shear angle $\theta$ was plotted
as a function of the magnetopause's $y$ and $z$ coordinates (it was
projected into the $x=0$ plane) and maxima in this shear angle map
were determined by finding maxima along cuts parallel to the
IMF~(K. Trattner, private communication, 2013).

The following sections detail how we robustly identify the
magnetopause and calculate model predictions in our global
magnetospheric simulations.

\subsection{Identification of the Magnetopause and Plasma Parameter
  Sampling}
\label{subsection::Magnetopause_Identification}
To identify the magnetopause in our global magnetospheric simulations,
we modify the method of~\citet{nemecek2011}.  (One may alternatively
adopt the method described in~\citet{palmroth2003} to identify the
magnetopause using streamlines, although this is not tested here.)
The current density magnitude is sampled radially from $7\le r< 20$
Earth radii $\left(\text{R}_\text{E}\right)$ at our highest grid cell
resolution (0.125~$\text{R}_\text{E}$). The location of maximum
current density magnitude at smallest $r$ is identified as the
magnetopause and this location's coordinates $\mathbf{r}_\text{MP}$
are saved; the chosen sampling range excludes the ring current, but we
take additional care by ensuring each point has a radial distance
within 2~$\text{R}_\text{E}$ of the previous point's radial distance.
We employ spherical coordinates in the Geocentric Solar Magnetospheric
(GSM) system, where the azimuthal angle $\phi$ is measured from the
$+x$ axis and the polar angle $\theta$ is measured from the $+z$ axis;
we discretize both into 5$^\circ$ increments. (Other increments were
tested and found to converge to the 5$^\circ$ results.)  We use an
azimuthal range of $-130^\circ\le\phi\le130^\circ$ to map the
magnetopause for northward IMF orientations
($\theta_\text{IMF}=\left[30^\circ,\,~60^\circ,\,~90^\circ\right]$ for
the purposes of this study) and $-110^\circ\le\phi\le110^\circ$ for
southward IMF orientations
($\theta_\text{IMF}=\left[120^\circ,\,~150^\circ,\,~165^\circ\right]$);
the polar angle has a range $0^\circ\le\theta\le180^\circ$ for both
northward and southward IMF orientations.

\begin{figure}[t]
\centering
\noindent\includegraphics[width=20pc]{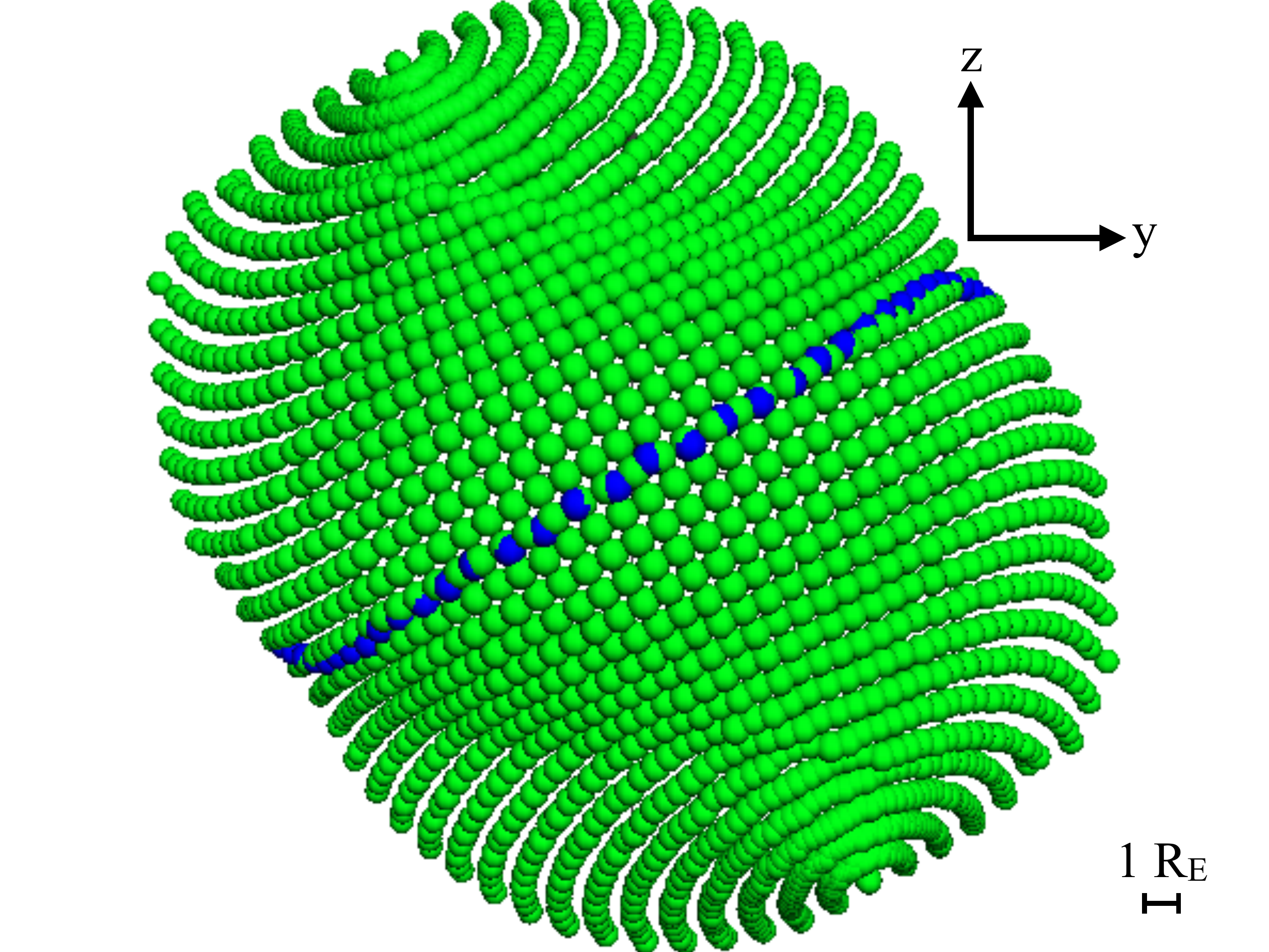}
\caption{Magnetopause locations (green) from maxima in the current
  density's magnitude in a simulation with
  $\theta_\text{IMF}=120^\circ.$ The magnetic separator (blue) is
  shown for reference.}
\label{fig::120_Magnetopause}
\end{figure}

An example of this process can be seen in
Fig.~\ref{fig::120_Magnetopause}, displaying the magnetopause surface
as green spheres for a simulation with IMF clock angle 120$^\circ$
that will be described in Section~\ref{subsection::Simulation_Study}.
Additionally, the separator for this simulation from~\citet{komar2013}
is plotted in blue.  It lies within the magnetopause surface, as it
should, confirming this method. The grid is rotated out of the
ecliptic plane; the spherical coordinate system is rotated
counter-clockwise about the $+x$~axis by an angle
$\theta_\text{IMF}/2$.  This is done to achieve near uniform sampling
near the separator and each model's prediction, {\it i.e.} near
$\theta\approx90^\circ$ in the rotated frame.

Once the approximate location of the magnetopause has been determined,
the local plasma parameters of the magnetosphere and magnetosheath are
sampled.  The appropriate way to do this is to sample along the
magnetopause normal $\mathbf{\hat{n}}$ at each magnetopause position
$\mathbf{r}_\text{MP}$.  One may perform a minimum variance analysis
(MVA) on the magnetic field~\citep{sonnerup1967} to determine the
magnetopause normal.  However, we find in practice this results in
incorrect magnetopause normals since the magnetospheric and
magnetosheath magnetic fields are parallel near the separator,
violating the core assumption of MVA.

Instead, the normal is calculated with the method described
in~\citet{hoppe1992}.  The algorithm is to take a single point on the
magnetopause $\mathbf{r}_\text{MP}$ and find its four nearest
neighbors $\mathbf{r}_k$ on the magnetopause, with
$k\in\left[1,\,2,\,3,\,4\right]$. Calculate the difference vector
$\mathbf{r}_k-\mathbf{r}_\text{MP}$, which gives the distance from the
point on the magnetopause and its $k$th nearest neighbor. Then,
construct the covariant matrix $\mathbf{M}$, whose $ij$ element is
\begin{equation}
  M_{ij}\left(\mathbf{r}_\text{MP}\right)=\sum_{k}\left(\mathbf{r}_k-\mathbf{r}_\text{MP}\right)_i\left(\mathbf{r}_k-\mathbf{r}_\text{MP}\right)_j,
\label{eqn::Normal_Covariance}
\end{equation}
where $i$ and $j$ refer to the $x$, $y$, or $z$ component in GSM
coordinates.  The eigenvector corresponding to the minimum eigenvalue
of $\mathbf{M}$ is the magnetopause normal; this normal direction is
chosen to point away from Earth, \emph{i.e.},
$\mathbf{\hat{x}}\cdot\mathbf{\hat{n}}\ge0$, and has been confirmed by
inspection.  This procedure has a physical interpretation of placing
several planes at $\mathbf{r}_\text{MP}$ with arbitrary orientations;
the normal of the plane that minimizes the distance between the plane
and the nearest neighbors is the magnetopause's normal.

The plasma parameters are sampled along the normal vector
$\mathbf{\hat{n}}$ at our highest resolution
(0.125~$\text{R}_\text{E}$) between $\mathbf{r}_\text{MP}-\left(5\
  ~\text{R}_\text{E}\right)\mathbf{\hat{n}}$ and
$\mathbf{r}_\text{MP}+\left(5\
  ~\text{R}_\text{E}\right)\mathbf{\hat{n}}$.  To accurately measure
the asymptotic plasma densities and magnetic fields undergoing
magnetic reconnection, we employ the method used
by~\citet{cassak2009c} for systems with asymmetries in either
parameter.  The current density is measured along $\mathbf{\hat{n}}$
to determine where it falls to $1/e$ of its maximum value in the
magnetosheath.  The distance between the current maximum and this
location is the current sheet's half-thickness $\delta$.  The upstream
plasma parameters of the magnetosphere are measured at
$\mathbf{r}_\text{MP}-\left(2\delta\right)\mathbf{\hat{n}}$ and those
of the magnetosheath at
$\mathbf{r}_\text{MP}+\left(2\delta\right)\mathbf{\hat{n}}$.  We
confirm the measured magnetospheric and magnetosheath magnetic fields
and densities are representative of the asymptotic values, as desired.
The raw upstream magnetic fields can have a small normal component
$\left(\left|B_n\right|\sim1\ ~\text{nT}\right)$; these are removed
prior to calculations.  These values are used as
$\mathbf{B}_\text{MS}$, $\mathbf{B}_\text{SH}$, $\rho_{MS}$, and
$\rho_{SH}$ for the models discussed in
Section~\ref{section::Reconnection_Location_Models}.

We note in passing that identifying the magnetopause as the maximum in
the current density's magnitude has its limitations.  One consequence
of asymmetric reconnection systems is that the stagnation point
separates from the reconnection X-line in 2D
reconnection~\citep{cassak2007c}; the maximum of the out-of-plane
current density also separates from the X-line.  We will describe in
Section~\ref{subsection::IMF_Strength_Test} a test of reconnection
location models in distinct simulations with IMF strengths of 20, 5,
and 2~nT. The separation distances between the current density maximum
and separator (the 3D version of the X-line) at the subsolar point in
these simulations increase as the IMF strength is lowered, with
distances of 0.07 (20~nT), 0.36 (5~nT), and
0.51~$\text{R}_\text{E}$~(2~nT); the separation distances are resolved
in the 5 and 2~nT simulations since our maximum resolution is
0.125~$\text{R}_\text{E}$.  In such systems, the current density
maximum does not coincide with the magnetopause.  For the purposes of
this study, we assume this feature corresponds to a compression of the
magnetopause in the $\mathbf{\hat{n}}$ direction at every point with
minimal impact to each model's calculation as the sampling method
described here measures the plasma parameters far enough away from the
magnetopause location.

\subsection{Finding Model Predictions Using Image Processing
  Techniques}
\label{subsection::Ridge_Detection}
Having found the approximate location of the magnetopause and the
parameters locally governing reconnection ($\rho_\text{MS}$,
$\rho_\text{SH}$, $\mathbf{B}_\text{MS}$, $\mathbf{B}_\text{SH}$), the
relevant quantities for each model are calculated at every point on
the magnetopause.  The magnetic shear angle $\theta$ is calculated
from Eq.~\eqref{eqn::Shear_Angle} at every point and is retained.
Then at every point, $\alpha$ is discretized in the range
$0<\alpha<\theta$.  For each value of $\alpha$, $B_{MS,L}$ and
$B_{SH,L}$ are calculated from Eqs.~\eqref{eqn::Magnetosphere_L}
and~\eqref{eqn::Magnetosheath_L} , and $c_{A,out}$, $E_{SP}$, and
$B_\text{MS,L}^2B_{SH,L}^2$ are calculated for that $\alpha$. The
$\alpha$ which maximizes the quantity in question is retained, and the
value of $c_{A,out}$, $E_{SP}$, and $B_\text{MS,L}^2B_{SH,L}^2$ is
retained for that point.  (One could also determine $\alpha$ from
$\partial c_{A,out}/\partial\alpha=0$ and $\partial
E/\partial\alpha=0$.) The current density magnitude is also retained
at every point.

\begin{figure}[t]
\centering
\noindent\includegraphics[width=20pc]{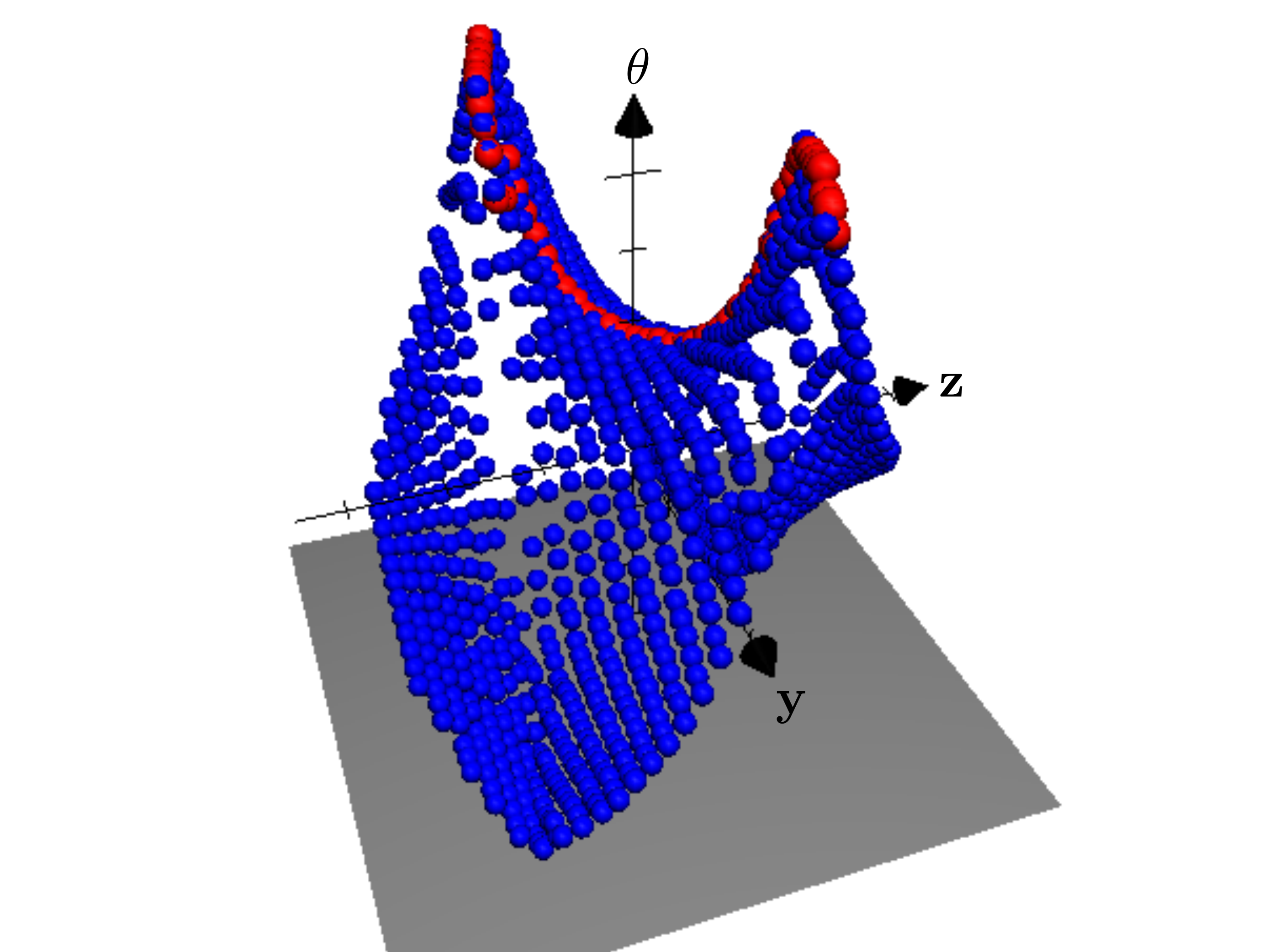}
\caption{Magnetic shear angle $\theta$ (blue spheres) as a function of
  the magnetopause's $y$ and $z$ coordinates and the corresponding
  ridge of maximum magnetic shear angle (red spheres) in a simulation
  with $\theta_\text{IMF}~=~120^\circ$.}
\label{fig::shear_surface}
\end{figure}

The results for each model are projected into the $x=0$ plane,
resulting in a 2D image $\mathscr{I}\left(y,\,z\right)$, where
$\mathscr{I}$ represents the quantity in question for each model.  An
example is shown in Fig.~\ref{fig::shear_surface} for the maximum
magnetic shear model, where $\mathscr{I}$ corresponds to $\theta$,
which is plotted as the third dimension in blue spheres for the
$\theta_\text{IMF}=120^\circ$ simulation.  The magnetic shear angle
surface has a saddle shape, as do the surfaces for all the other
models tested (not shown).  This plot makes it clear that finding the
collection of locations where the quantity in question is locally a
maximum is tantamount to finding the ridge of local maxima in
$\mathscr{I}$.

We achieve this in a robust way using image processing techniques
using a ridge finding algorithm~\citep{lindeberg1993,lindeberg1998}.
First, we construct the Hessian tensor
$H_{ij}\left(y,\,z\right)=\partial^2\mathscr{I}\left(y,\,z\right)/\partial
x_i\partial x_j$ via finite differencing at every point in the image.
Diagonalizing the Hessian matrix defines a $\left(p,\,q\right)$
coordinate system, where the eigenvector associated with the maximum
absolute eigenvalue points in the direction of the local
maxima. (Without loss of generality, we designate that eigenvector as
$\mathbf{\hat{p}}$, the unit vector for the $p$ direction.)  Local
maxima are determined by finding all locations where $\partial
\mathscr{I}/\partial p=0$ (local extrema test) and $\partial^2
\mathscr{I}/\partial p^2<0$ (local maxima test) using linear
interpolation between nearest neighbors~\citep{lindeberg1993}.

The results of this technique for the maximum magnetic shear model for
$\theta_\text{IMF}=120^\circ$ are shown as the red spheres in
Fig.~\ref{fig::shear_surface}. These red spheres clearly mark the
ridge of the magnetic shear surface as desired.  The methodology we
employ yields qualitatively similar results to previous studies
testing the maximum magnetic shear model for comparable magnetospheric
and solar wind conditions~(see Fig.~1 in~\citet{fuselier2011}),
indicating this is an appropriate technique to determine each model's
prediction. A more thorough comparison between these techniques will
be presented in Section~\ref{subsection::Clock_Angle_Dependence}.

\subsection{Determining Magnetic Separators}
\label{subsection::Determing_Magnetic_Separators}
We employ the separator mapping algorithm of~\citet{komar2013} which
has been shown to reliably trace the dayside portion of the magnetic
separators connecting the magnetic nulls in global magnetosphere
simulations for any IMF direction.  When the IMF has a northward
component, observations have shown that reconnection occurs poleward
of the magnetic cusps (see~\citet{wilder2014}, and references
therein).  We therefore trace portions of the magnetic separator
nightward of the nulls in our simulations with
$\theta_\text{IMF}=30^\circ$, 60$^\circ$, and 90$^\circ$.

In the separator tracing algorithm, a hemisphere is initially centered
around a magnetic null.  The hemisphere's surface, of radius
1~$\text{R}_\text{E}$ for our purposes, is discretized into a grid.
The magnetic field lines piercing the hemisphere at each grid point
are traced to determine their magnetic topology: closed terrestrial,
solar wind, and open either connected to the north or south magnetic
poles.  The approximate location of the separator is determined by
finding where these four magnetic topologies meet on a hemisphere's
surface.  Then, another hemisphere is centered at the determined
separator location, and the procedure is iterated to trace the
separator.  The dayside separator is traced from northern to southern
null in this fashion and the nightside portion is traced back for 10
$\text{R}_\text{E}$ from the magnetic nulls when the IMF has a
northward orientation.

\subsection{Magnetospheric Simulation Study}
\label{subsection::Simulation_Study}
We perform global simulations using the Block Adaptive Tree Solarwind
Roe-type Upwind Scheme (BATS-R-US)
code~\citep{powell1999,gombosi2000,dezeeuw2000,toth2012a}, although
the methods detailed in the previous sections can be adapted to other
global magnetospheric codes.  Simulations are performed at NASA's
Community Coordinated Modeling Center (CCMC) and are partially
analyzed with CCMC's Kameleon software suite.  The simulation domain
is $-255<x<33$, $-48<y<48$, and $-48<z<48$, where distances are
measured in $\text{R}_\text{E}$ and the coordinate system is GSM.

The simulations are carried out using BATS-R-US version 8.01. The
simulations are evolved for two hours (02:00:00) of magnetospheric
time. We look at the 02:00:00 mark of simulation data as the system
has achieved a quasi-steady state by this time. The standard
high-resolution grid for CCMC simulations has $1,958,688$ grid cells
with a coarse resolution of $8\ ~\text{R}_\text{E}$ in the far
magnetotail, and a fine resolution of $0.25\ ~\text{R}_\text{E}$ near
the magnetopause. The present study employs a higher resolution grid
of 0.125~$\text{R}_\text{E}$ packed in the region $-15<
x,\,y,\,z<15$~$\text{R}_\text{E}$, totaling $16,286,400$ simulation
grid cells.

The initial simulations do not employ a dipole tilt, although we
include one later.  All simulations use fixed solar wind inflow
conditions. The solar wind has temperature $T=232,100\ ~\text{K}$, IMF
strength $20\ ~\text{nT}$, number density $n=20\ ~\text{cm}^{-3}$, and
a solar wind velocity of ${\bf v}=-400\ ~\text{km/s}\
~\mathbf{\hat{x}}$. We perform distinct simulations with IMF clock
angles $\theta_\text{IMF}=30^\circ$, 60$^\circ$, 90$^\circ$,
120$^\circ$, 150$^\circ$, and 165$^\circ$. The IMF does not have a
$B_x$ component. Constant Pederson and Hall conductances of 5 mhos are
used. The solar radio flux F10.7 index is set at a value of 150.

The parameter regime of the solar wind in the present global
magnetospheric simulations employed atypical IMF strengths and solar
wind densities; both quantities are an order of magnitude higher than
typically quiet solar wind conditions.  This choice is made so the
dayside magnetosphere is smaller, and the region of high resolution
need not be as large. To ensure that the results are reliable, it is
important to check that the polar cap is not saturated with these
solar wind parameters~\citep{lopez2010}.

We measure the polar cap potential difference in three distinct
simulations with $B_\text{IMF}=2$, 5, and 20~nT.  These simulations
have no dipole tilt, the IMF clock angle is fixed at $120^\circ$, and
all other solar wind parameters are the same as before.
\begin{figure}[t]
\centering
\noindent\includegraphics[width=20pc]{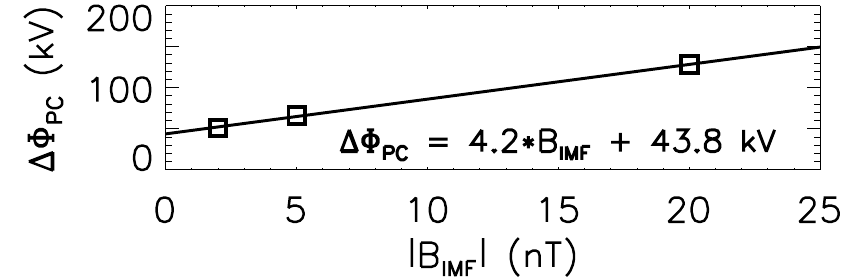}
\caption{Cross polar cap potential difference $\Delta\Phi_\text{PC}$
  in simulations with $\left|\mathbf{B}_\text{IMF}\right|=2,\,5$ and
  20~nT for $\theta_\text{IMF}~=~120^\circ$.  The potential difference
  is linear in $\left|\mathbf{B}_\text{IMF}\right|$, so the polar cap
  is not saturated.}
\label{fig::PC_Saturation}
\end{figure}
Figure~\ref{fig::PC_Saturation} displays the measured polar cap
potential difference as a function of
$\left|\mathbf{B}_\text{IMF}\right|$.  The plot clearly reveals that
the potential difference remains linear as a function of IMF strength,
implying the polar cap is not saturated.  For
comparison,~\citet{lopez2010}, using LFM simulations, showed that the
polar cap potential difference remains linear for
$B_\text{IMF}<11.25$~nT with solar wind density
$n_\text{SW}=5$~cm$^{-3}$ and for $B_\text{IMF}<15$~nT when
$n_\text{SW}=8$~cm$^{-3}$ using a solar wind speed of 400~km/s and
ionospheric Pedersen conductance of 5~mhos.  As our simulations employ
$n_\text{SW}=20$~cm$^{-3}$, it is reasonable that we find the polar
cap is not saturated for $B_\text{IMF}=20$~nT. Our measured viscous
potential (the $y$-intercept) is 43.8~kV, consistent with the trend
seen by~\citet{lopez2010}, with their measured viscous potentials of
33.9~and~41.9~kV for $n_\text{SW}=5$~and~8~cm$^{-3}$, respectively.
We are therefore confident that the polar cap is not saturated for our
simulation parameters.

For the present simulations, we employ a uniform explicit resistivity
$\eta$.  It is known that Earth's magnetopause is collisionless, but
including an explicit resistivity allows for reproducible results that
are independent of the numerics.  With the exception of the asymmetric
Sweet-Parker reconnection rate
[Eq.~\eqref{eqn::SP_Asymmetric_Reconnection_Rate}], all of the models
selected for testing are independent of the reconnection dissipation
mechanism.  Therefore, the model predictions should likewise remain
independent of this choice for \emph{any} self-consistent
magnetospheric model.  We include an explicit resistivity
$\eta/\mu_0=6.0~\times~10^{10}$~m$^2$/s in our simulations, which
allows us to sufficiently resolve the dayside magnetopause to
determine the validity of the models tested.  We refer the reader
to~\citet{komar2013} for a more detailed discussion on the inclusion
of an explicit resistivity.
\begin{figure*}
\centering
\noindent\includegraphics[width=39pc]{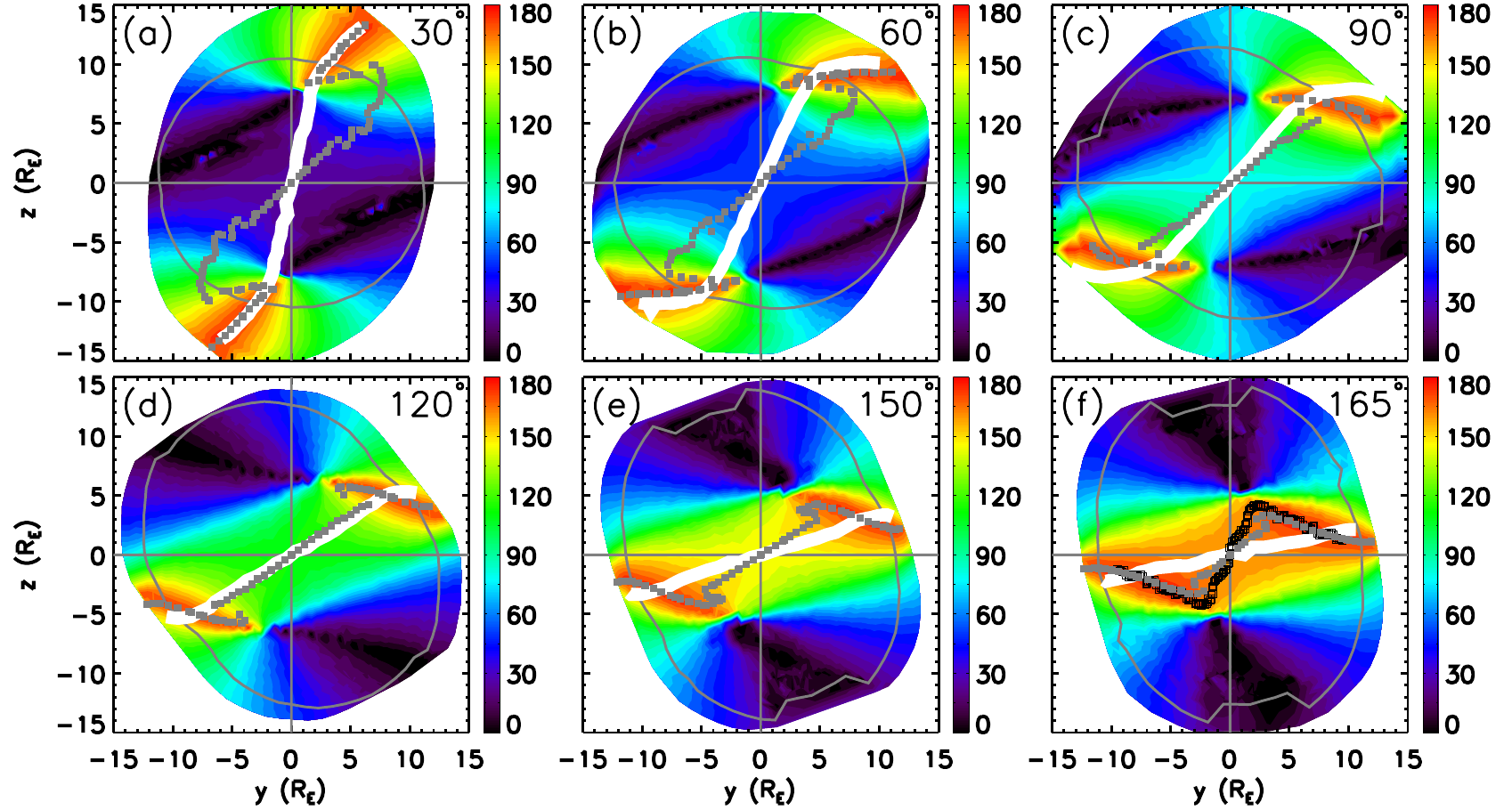}
\caption{Results for the maximum magnetic shear model in simulations
  with $\theta_\text{IMF}$ of (a) 30$^\circ$, (b) $60^\circ$, (c)
  $90^\circ$, (d) $120^\circ$, (e) $150^\circ$, and (f) $165^\circ$.
  The calculated magnetic shear angle $\theta$ is the color
  background, the gray squares display the curve of maximum magnetic
  shear angle, and the magnetic separator is displayed in white.  The
  gray oval displays the magnetopause's projection in the $x=0$ plane.
  Maxima determined from cuts along the IMF
  direction~\citep{trattner2007} are displayed as black squares in
  panel (f).}
\label{fig::Magnetic_Shear}
\end{figure*}

\begin{figure*}
\centering
\noindent\includegraphics[width=39pc]{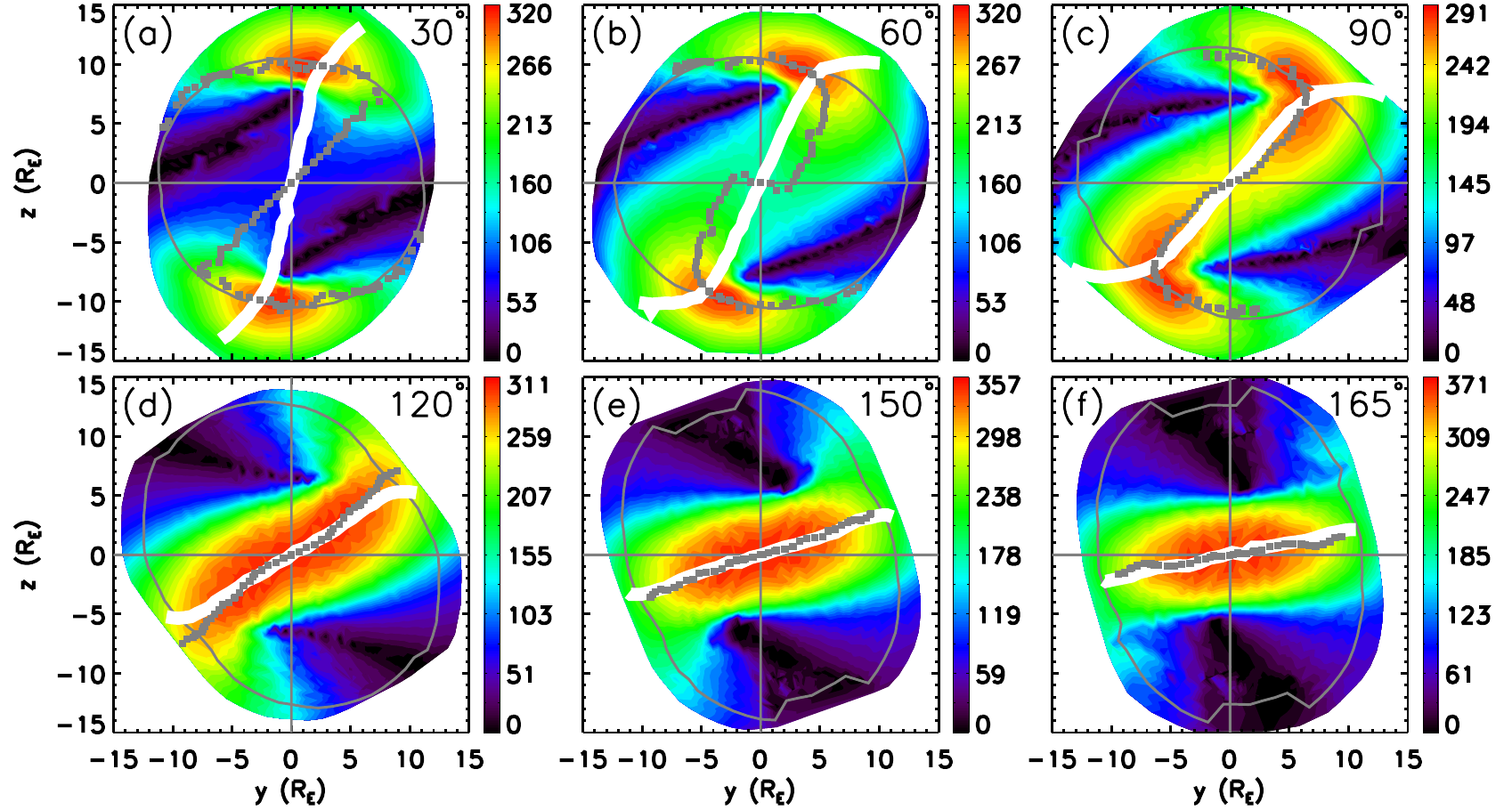}
\caption{Results for the asymmetric outflow speed $c_{A,out}$.  See
  the caption of Figure~\ref{fig::Magnetic_Shear} for definitions.}
\label{fig::Outflow_Speed}
\end{figure*}

\subsection{Simulation Resolution}
\label{subsection::Simulation_Resolution}
The numerical grid described in the present study is different from
those of~\citet{komar2013} where the algorithm for finding separators
was presented.  The present simulations employed a resolution of 0.125
$\text{R}_\text{E}$ in the region
$-15~<~x,\,y,\,z~<~15$~$\text{R}_\text{E}$ whereas~\citet{komar2013}
used a resolution of 0.125~$\text{R}_\text{E}$ in the region
$-6<x<10$, $-10<y<10$, $-5<z<5$~$\text{R}_\text{E}$, totaling a
difference of $\approx$12.6~million simulation grid cells.  The
present study uses the larger high-resolution domain to ensure that
the entire dayside magnetopause (and therefore the magnetic
separators) is within the high-resolution domain.  The magnetic
separators in the present simulations are not significantly different
than those in~\citet{komar2013}.

\section{Results}
\label{section::Results}
\begin{figure*}
\centering
\noindent\includegraphics[width=39pc]{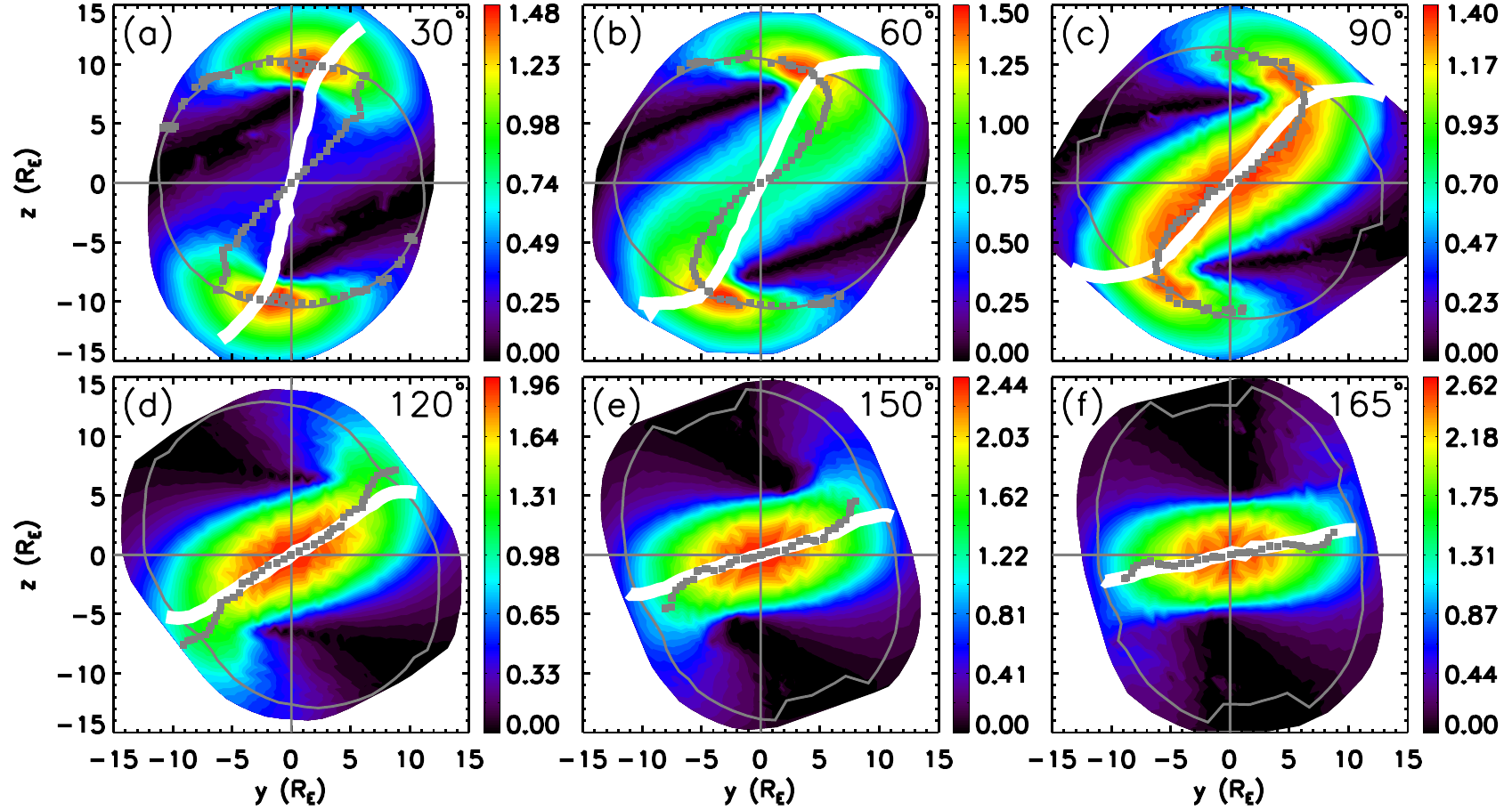}
\caption{Results for the asymmetric Sweet-Parker reconnection rate
  $E_{SP}$.  See the caption of Figure~\ref{fig::Magnetic_Shear} for
  definitions.}
\label{fig::Reconnection_Rate}
\end{figure*}

\begin{figure*}
\centering
\noindent\includegraphics[width=39pc]{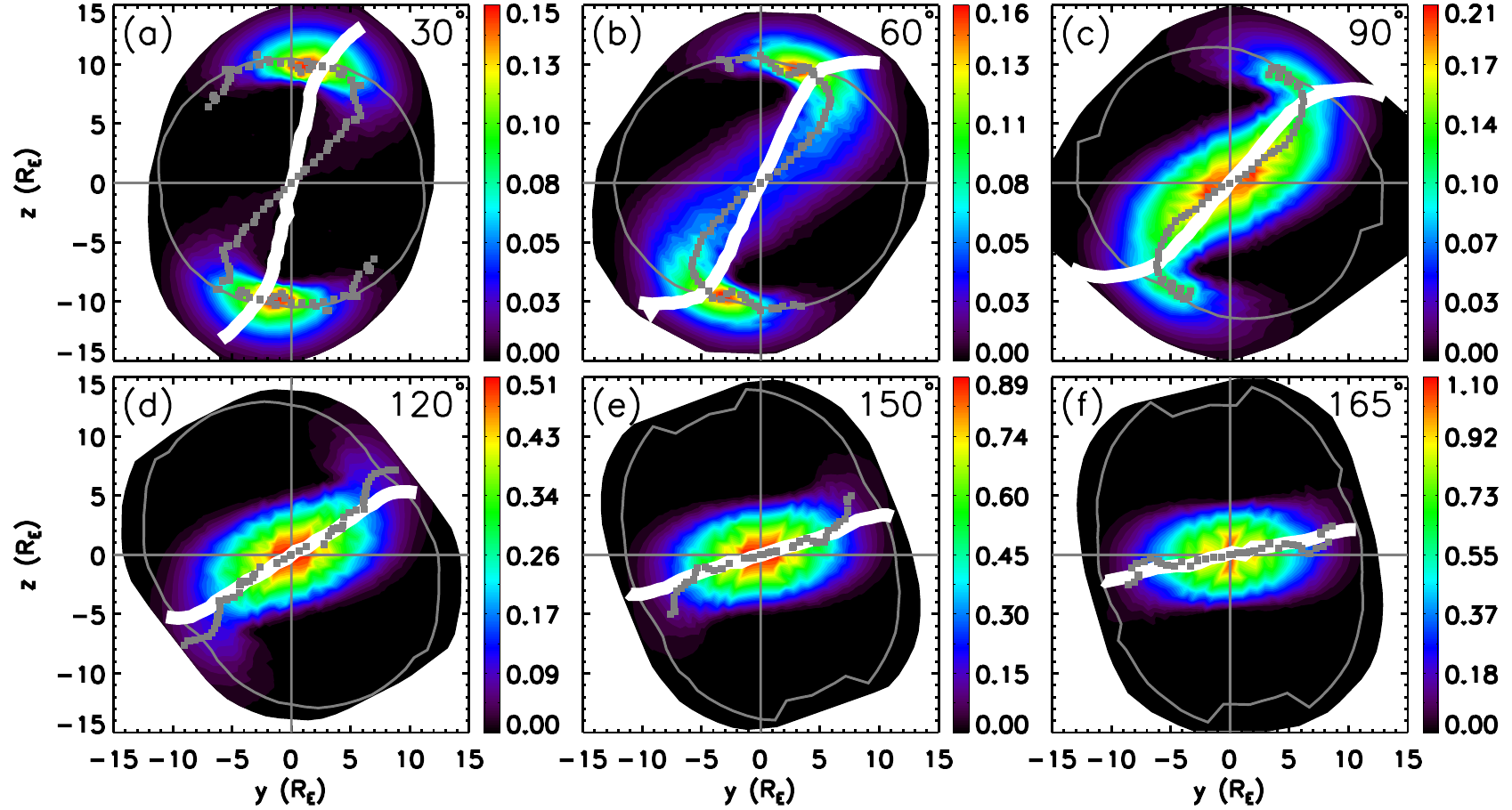}
\caption{Results for the angle of bisection.  See the caption of
  Figure~\ref{fig::Magnetic_Shear} for definitions.}
  \label{fig::Bisection_Angle}
\end{figure*}
We present results for the predicted locations for the models
discussed in Section~\ref{section::Reconnection_Location_Models} and
compare the results to magnetic separators.
Figures~\ref{fig::Magnetic_Shear}-\ref{fig::Current_Density} display
the results in a common format.  The quantity of interest is displayed
as the background color, the model's prediction is displayed with
solid gray squares, and the magnetic separators determined with the
algorithm described in~\citet{komar2013} are displayed as the solid
white line.  The magnetopause's intersection with the $x=0$ plane is
displayed as the gray oval. The panels for each figure correspond to
IMF clock angles~$\theta_\text{IMF}$ of (a)~30$^\circ$,
(b)~60$^\circ$, (c)~90$^\circ$, (d)~120$^\circ$, (e)~150$^\circ$, and
(f)~165$^\circ$.  Figure~\ref{fig::Magnetic_Shear} displays the
magnetic shear angle $\theta$ in degrees calculated from
Eq.~\eqref{eqn::Shear_Angle}, Fig.~\ref{fig::Outflow_Speed} has the
asymmetric outflow speed $c_{A,out}$ in km/s calculated from
Eq.~\eqref{eqn::Asymmetric_Outflow_Speed},
Fig.~\ref{fig::Reconnection_Rate} has the asymmetric Sweet-Parker
reconnection rate $E_{SP}$ in mV/m calculated from
Eq.~\eqref{eqn::SP_Asymmetric_Reconnection_Rate},
Fig.~\ref{fig::Bisection_Angle} has the quantity
$B_\text{MS,L}^2B_\text{SH,L}^2$ from~Eq.~\eqref{eqn::E_Hesse} in
10$^8$~nT$^4$, and Figure~\ref{fig::Current_Density} has the current
density magnitude $\left|\mathbf{J}\right|$ in nA/m$^2$.

Figure~\ref{fig::Uniform_Guide_Field} displays the results for the
uniform $B_M$ component model.  Plots (a) through (f) are for the same
IMF clock angle simulations as previously discussed, with a few
notable differences.  The local $B_M$ component (guide field) is
calculated by determining the orientation angle $\alpha$ from
Eq.~\eqref{eqn::Sonnerup_Criterion}, and rotating about the determined
boundary normal $\mathbf{\hat{n}}$.  The gray arrows in
Fig.~\ref{fig::Uniform_Guide_Field} display the projection of the
magnetic field's $M$ component in the $x=0$ plane; locations without
any vector displayed are where reconnection is predicted to be
geometrically impossible.  The magnetic separator is displayed as the
solid black line, with the magnetopause's location in the $x=0$ plane
plotted in gray as before.

Section~\ref{subsection::Clock_Angle_Dependence} discusses how the
predictions of each model change with IMF clock angle for fixed solar
wind conditions.  We test the dependence on IMF strength of various
models in Section~\ref{subsection::IMF_Strength_Test}.
Section~\ref{subsection::Dipole_Tilt_Dependence} describes results
from two distinct simulations that include a dipole tilt with IMF
clock angles 120$^\circ$ and 30$^\circ$.  Lastly,
Section~\ref{subsection::Shear_Flow} discusses an attempt to
incorporate the effects of a plasma flow shear on reconnection into
the outflow speed and reconnection rate models from
Section~\ref{section::Reconnection_Location_Models}.

\subsection{Clock Angle Dependence}
\label{subsection::Clock_Angle_Dependence}
Before assessing each model, we perform a more careful comparison
between our ridge finding technique and the one
in~\citet{trattner2007}.  We plot the result using the technique
of~\citet{trattner2007}, which finds maxima of magnetic shear angle
along cuts parallel to the IMF direction, as black squares in
Fig.~\ref{fig::Magnetic_Shear}(f) for the
$\theta_\text{IMF}=165^\circ$ simulation.  While qualitatively
similar, there are apparent differences.  If one takes cuts along
different directions (not shown), the detected locations change, and
can move to locations on the other side of the image processing ridge.
This underscores two points: (1) it is important to employ a
user-independent technique, such as image processing, to locate each
models' prediction as the~\citet{trattner2007} algorithm exhibits
sensitivity to the cut orientation and (2) the noticeable changes in
the prediction are caused by the relative flatness in the saddle
region that occurs near the subsolar point in these results.
Consequently, there are uncertainties that could cause minor changes
to the predictions which should be kept in mind in the interpretation
of these results.

We now focus on the predictions of each model and how they compare
with the separators for all IMF clock angles. Of the models tested,
the maximum magnetic shear as shown has the largest discrepancy with
the dayside portion of the magnetic separators in simulations without
a dipole tilt, but could have better agreement in light of the
uncertainties discussed above.  More importantly, only the maximum
magnetic shear model reproduces the portion of the magnetic separator
nightward of the magnetic nulls with high fidelity.  These portions of
the magnetic separators coincide with the antiparallel regions and
are near the magnetic cusps, consistent with early
expectations~\citep{dungey1961,dungey1963,crooker1979} and
with observations~(see~\citet{wilder2014}, and references
therein). 

The asymmetric reconnection arguments ($c_{A,out}$ and $E_{SP}$) show
better agreement with the magnetic separators overall for the
simulations with southward IMF orientations.  The ridges of these
models rotate around the $x$ axis, as does the separator, although
there are differences between the separators and the model predictions
for clock angles $\theta_\text{IMF}\le120^\circ$.

\begin{figure*}
\centering
\noindent\includegraphics[width=39pc]{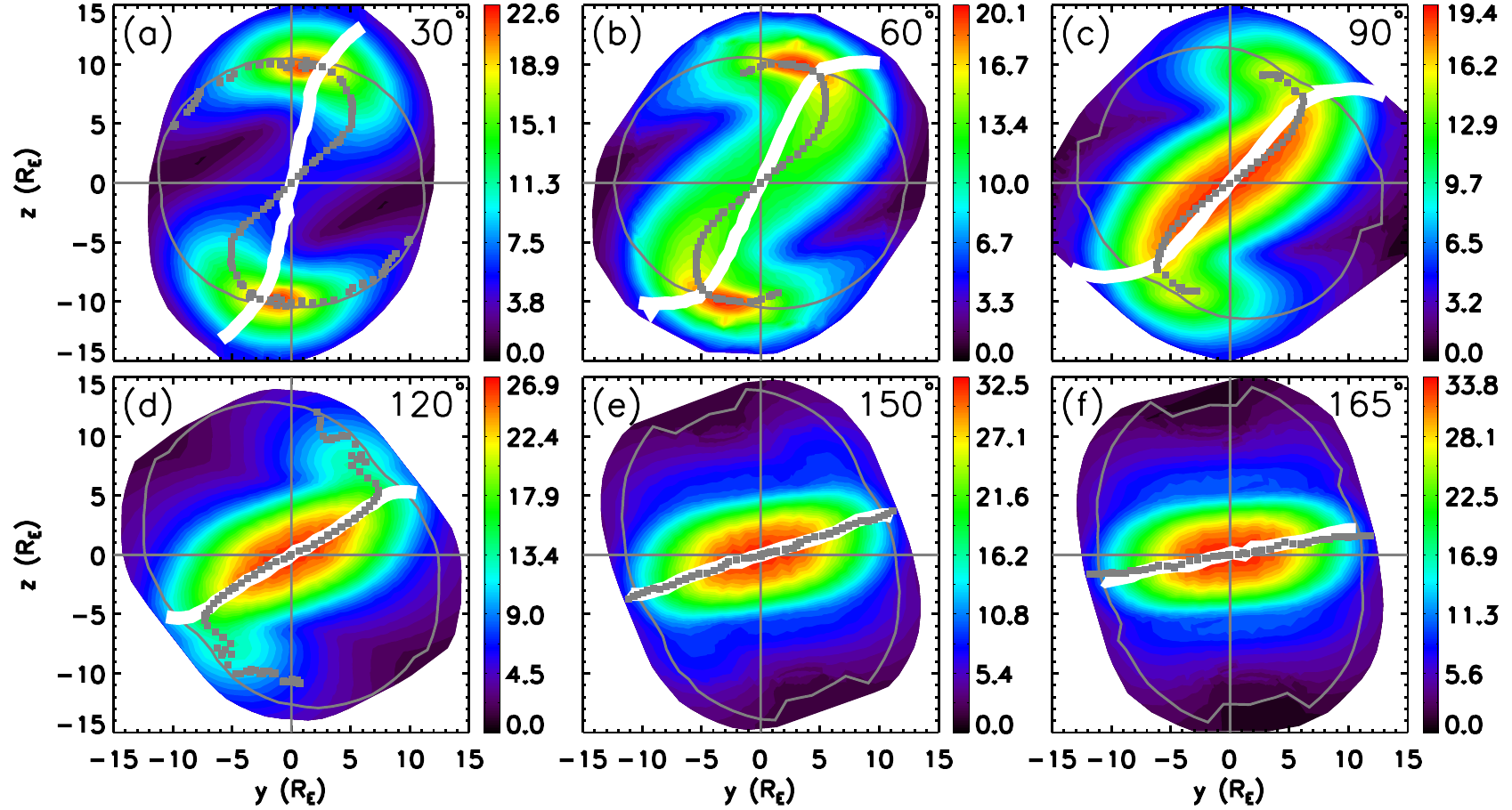}
\caption{Results for the current density magnitude
  $\left|\mathbf{J}\right|$.  See the caption of
  Figure~\ref{fig::Magnetic_Shear} for definitions.}
\label{fig::Current_Density}
\end{figure*}

\begin{figure*}
\centering
\noindent\includegraphics[width=39pc]{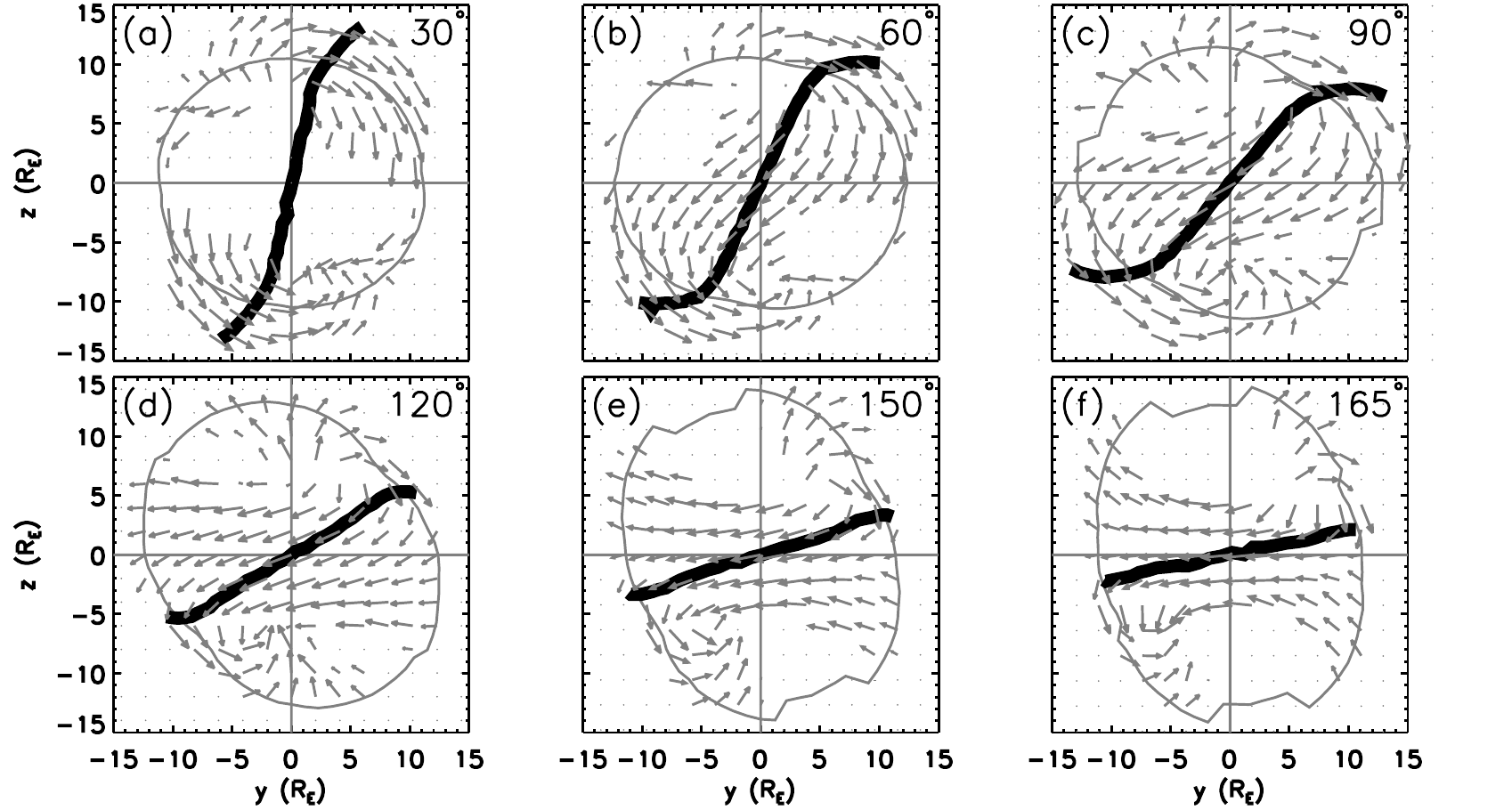}
\caption{Results for the uniform $B_M$ component model.  Panels (a)
  through (f) are from the same simulations as in
  Figures~\ref{fig::Magnetic_Shear}-\ref{fig::Current_Density}.  The
  $y$ and $z$ components of $B_M$ are plotted as gray arrows and the
  magnetic separator is displayed in black.}
\label{fig::Uniform_Guide_Field}
\end{figure*}

The predictions for the angle of bisection and the maximum current
density models have a similar shape as the other models' predictions
through the subsolar point.  Figures~\ref{fig::Bisection_Angle}(a)-(c)
and~\ref{fig::Current_Density}(a)-(c) show intersections between the
model predictions and magnetic separators near the antiparallel
regions and at the subsolar point for $\theta_\text{IMF}\le90^\circ$.
The predictions of these arguments do follow the separator more
closely for southward IMF orientations, and do not rotate around the
magnetopause for northward orientations.

The uniform $B_M$ component model would accurately predict the
separator if the arrows point along the separator.  They clearly do
not; at the subsolar point, the measured orientation angle $\alpha$
relative to the magnetospheric magnetic field is~(see
Fig.~\ref{fig::LMN_Coordinates}): (a) N/A (reconnection is
geometrically impossible), (b) 46.9$^\circ$, (c) 53.1$^\circ$, (d)
64.1$^\circ$, (e) 76.8$^\circ$, and (f) 83.6$^\circ$.  These angles
are inconsistent with the orientation angles $\alpha_\text{Sep}$ of
the magnetic separators (a) 10.3$^\circ$, (b) 21.7$^\circ$, (c)
35.8$^\circ$, (d) 55.3$^\circ$, (e) 71.2$^\circ$, and (f)
77.5$^\circ$.  The discrepancies between the magnetic field's $M$
component and the orientation of the magnetic separators provide
further evidence that this description does not accurately predict the
orientation of magnetic reconnection, consistent with previous
observations~\citep{teh2008} and simulations~\citep{schreier2010}.

\subsection{IMF Strength Dependence}
\label{subsection::IMF_Strength_Test}

Equations~\eqref{eqn::Asymmetric_Outflow_Speed}
and~\eqref{eqn::SP_Asymmetric_Reconnection_Rate} have strong
dependencies on the magnetospheric and magnetosheath magnetic field
strengths.  The magnetospheric magnetic field strength is usually much
larger than that of the magnetosheath, and does not vary much for the
chosen simulation parameters.  This implies that the reconnection
efficiency models depend strongly on the magnetosheath magnetic field
strength $B_\text{SH}$.  Meanwhile, the maximum magnetic shear model
is expected to be independent of $B_\text{SH}$.  We therefore vary
$B_\text{SH}$ to see if it helps to distinguish among these models.

\begin{figure*}
\centering
\noindent\includegraphics[width=39pc]{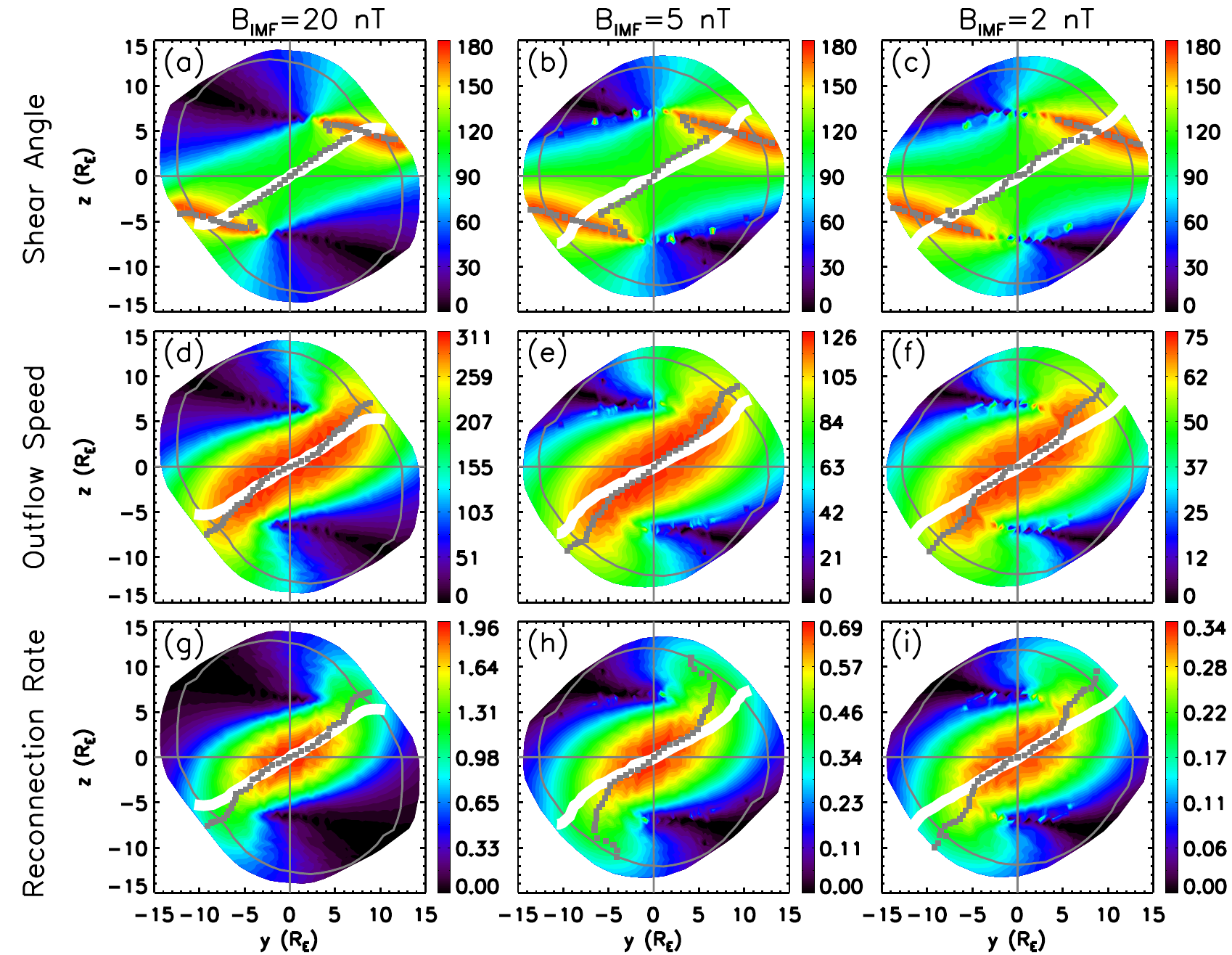}
\caption{Results for testing model dependence on IMF strength for:
  (top row) the magnetic shear angle, (middle row) asymmetric outflow
  speed, and (bottom row) the asymmetric Sweet-Parker reconnection
  rate. Plots (a), (d) and (g) are for $B_\text{IMF}=20$~nT, (b), (e),
  (h) are for $B_\text{IMF}=5$~nT, and (c), (f), (i) for
  $B_\text{IMF}=2$~nT.  See the caption of
  Figure~\ref{fig::Magnetic_Shear} for definitions.}
  \label{fig::IMF_Strength_Test}
\end{figure*}

We lower the IMF strength from 20~nT to 5 and 2~nT in separate
simulations and use $\theta_\text{IMF}=120^\circ$ with all other
simulation parameters unchanged.  Figure~\ref{fig::IMF_Strength_Test}
displays the results from these simulations.  The top row displays the
magnetic shear angle in degrees, the middle row has the asymmetric
outflow speed in km/s, and the bottom row has the asymmetric
Sweet-Parker reconnection rate in mV/m.
Figures~\ref{fig::IMF_Strength_Test}(a), (d), and (g) are for the
$B_\text{IMF}=20$~nT simulation, (b), (e), and (h) have
$B_\text{IMF}=5$~nT, and (c), (f), and (i) use $B_\text{IMF}=2$~nT.
As before, each model's prediction is plotted as the gray squares, the
magnetic separators are displayed in white.
 
We note in passing that the magnetopause is elongated along the IMF
direction in the 20~nT plots of Fig.~\ref{fig::IMF_Strength_Test}.
This effect is consistent with observations~\citep{lavraud2013} and
simulations~\citep{lavraud2008} for low solar wind Alfv\'en Mach
number.

The subsolar region in each simulation becomes increasingly asymmetric
as the IMF strength is lowered; the measured magnetic field ratios are
$b=B_\text{MS}/B_\text{SH}=1.4$, 4.1, and 11 for $B_\text{IMF}=20$, 5,
and 2~nT, respectively.  As seen in Fig.~\ref{fig::IMF_Strength_Test},
each model's prediction becomes noisier for smaller $B_\text{SH}$ due
to the limitation of determining the magnetopause via the current
maximum, as described in
section~\ref{subsection::Magnetopause_Identification}.

Figures~\ref{fig::IMF_Strength_Test}(a)-(c) show the maximum magnetic
shear model remains relatively independent of the IMF strength as
expected; this model's predictions have very similar structures as
seen previously for arbitrary $\theta_\text{IMF}$, making similar
angles with the magnetospheric magnetic field of 49.6$^\circ$,
50.2$^\circ$, and 53.2$^\circ$ in simulations with $B_\text{IMF}=20$,
5, and 2~nT, respectively.  This model maps large portions of the
magnetic separator across the dayside magnetopause, consistent with
previous results.

It is difficult to see differences between the asymmetric outflow
speed and asymmetric Sweet-Parker reconnection rate models in
Fig.~\ref{fig::IMF_Strength_Test}. The predictions of both models
closely map the magnetic separator in these simulations.  For a more
quantitative comparison between these models, we calculate each
model's predicted orientation angle and compare these values to the
separator's orientation at the subsolar point.  The separator's
orientation angle $\alpha_{Sep}$ measured from the magnetospheric
magnetic field is 55.3$^\circ$, 61.5$^\circ$, and 56.2$^\circ$ in the
$B_\text{IMF}=20$, 5, and 2~nT simulations, respectively.  The
orientation of the ridges passing through the subsolar region for the
asymmetric outflow speed make an angle $\alpha_{V}$ and, likewise, the
asymmetric Sweet-Parker reconnection rate $\alpha_E$ are again
measured from the magnetospheric magnetic field, with values
$\alpha_{V}=47.0^\circ$ and $\alpha_E=51.0^\circ$ in the
$B_\text{IMF}=20$~nT simulation, $\alpha_{V}=45.7^\circ$ and
$\alpha_E=49.7^\circ$ for $B_\text{IMF}=5$~nT, and finally,
$\alpha_{V}=47.4^\circ$ and $\alpha_E=47.4^\circ$ with
$B_\text{IMF}=2$~nT.

The orientation angle $\alpha_E$ decreases as IMF strength is lowered,
while $\alpha_V$ displays a minimum when $B_\text{IMF}=5$~nT.
However, $\alpha_{Sep}$ is maximum when $B_\text{IMF}=5$~nT.  All of
the models tested have predictions that are reasonably close, but none
show perfect agreement with the orientation at the subsolar
point. This implies that these models will need further modification
to accurately predict the orientation of reconnection at Earth's
subsolar magnetopause.

\subsection{Dipole Tilt Dependence}
\label{subsection::Dipole_Tilt_Dependence}
Including a dipole tilt can alter the location of magnetic
reconnection.  For example,~\citet{trattner2007} showed that the
prediction of the maximum magnetic shear model moves southward under
positive dipole tilt (as measured from the $+z$ axis in the $x$-$z$
GSM plane, \textit{i.e.}~the northern magnetic pole pointing Sunward),
and likewise moves northward for negative dipole tilt when paired with
southward IMF orientations.  Reconnection in simulations has also been
measured to change location in simulations with a dipole
tilt~\citep{park2006,palmroth2012,cnossen2012,hoilijoki2014}.

Employing a more realistic dipole model of Earth may assist in
distinguishing among the models.  Note in
Figs.~\ref{fig::Magnetic_Shear}-\ref{fig::Current_Density} the model
predictions intersect the magnetic separators at the subsolar point
and near the anti-parallel region for all IMF clock angles.  These
intersections result from symmetry; including a dipole tilt will break
this symmetry making it easier to distinguish which model predicts the
magnetic separator.

\begin{figure*}
\centering
\noindent\includegraphics[width=39pc]{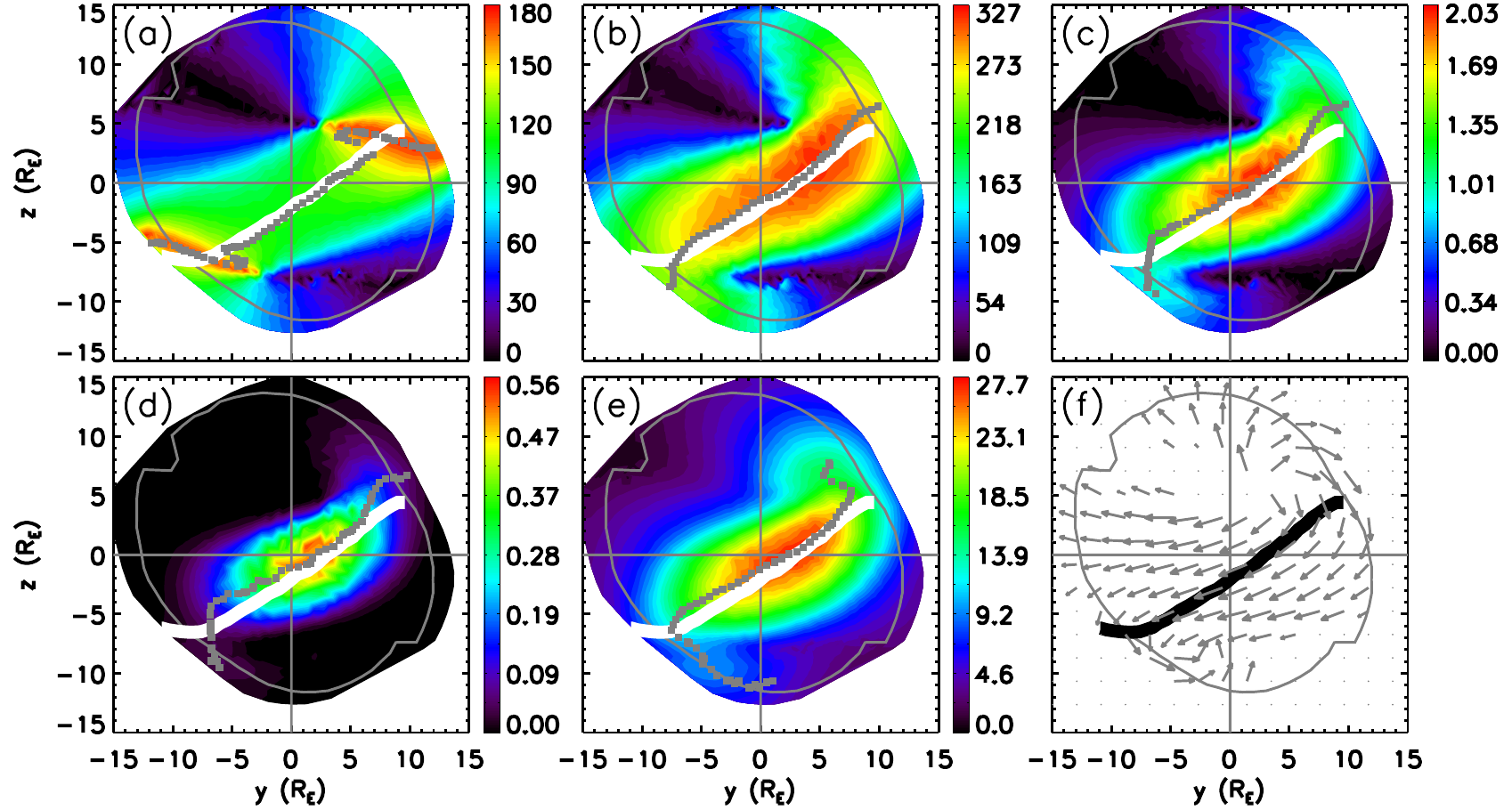}
\caption{Results from a simulation with a dipole tilt of $+15^\circ$
  and a southward IMF orientation with clock angle 120$^\circ$: (a)
  the magnetic shear angle $\theta$, (b) the asymmetric outflow speed
  $c_{A,out}$, (c) the asymmetric Sweet-Parker reconnection rate
  $E_{SP}$, (d) the angle of bisection, (e) the current density
  magnitude $\left|\mathbf{J}\right|$, and (e) the $y$ and $z$
  components of the uniform $B_M$ component.  See the caption of
  Figure~\ref{fig::Magnetic_Shear} for definitions.}
\label{fig::120_Dipole_Tilt}
\end{figure*}

\begin{figure*}
\centering
\noindent\includegraphics[width=39pc]{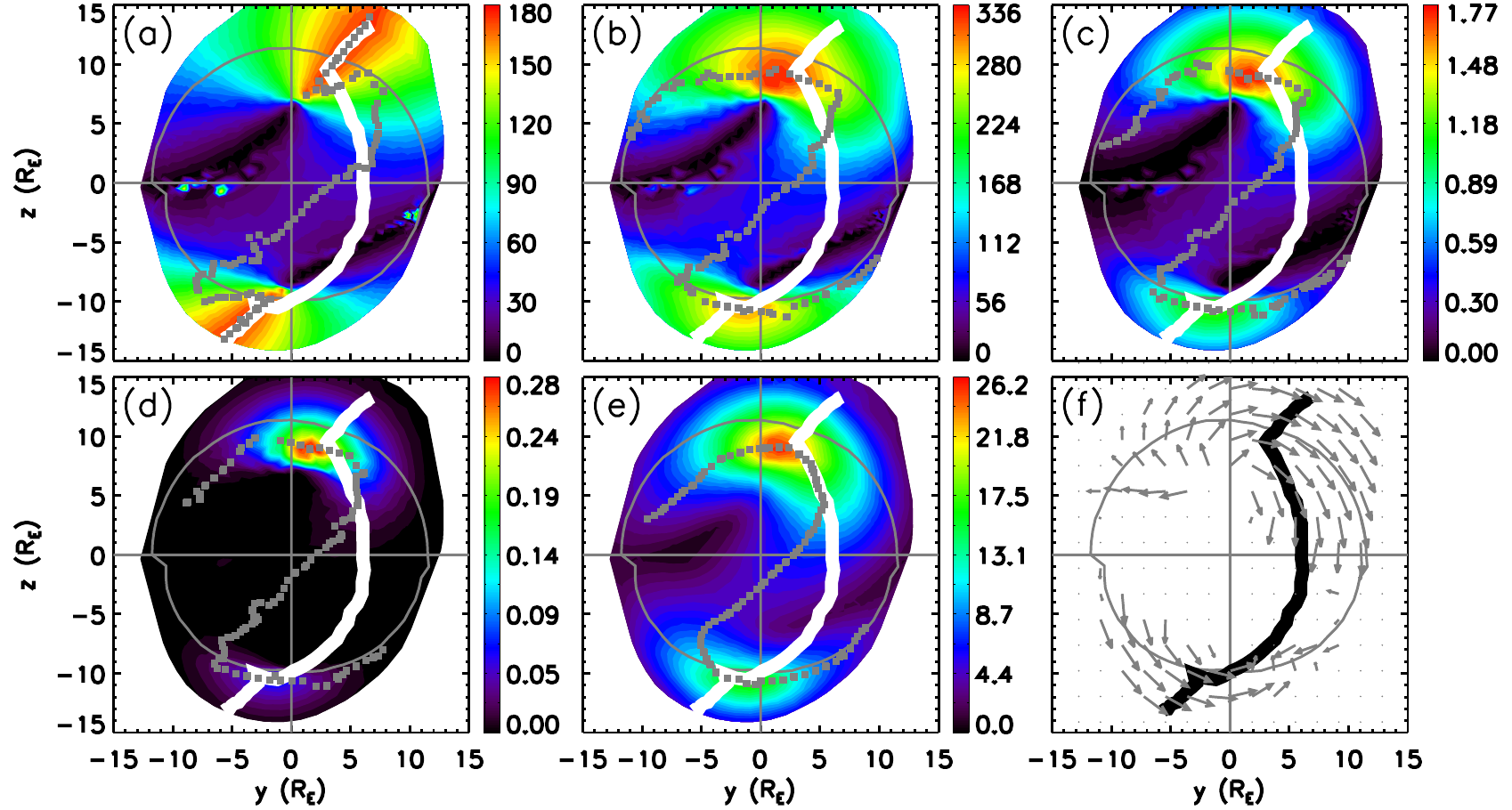}
\caption{Results from a simulation with dipole tilt of $+15^\circ$ and
  a northward IMF orientation with clock angle 30$^\circ$. See the
  caption of Figure~\ref{fig::120_Dipole_Tilt} for definitions.}
\label{fig::30_Dipole_Tilt}
\end{figure*}

We present the results of two additional simulations that employ a
$+15^\circ$ dipole tilt.  The same solar wind parameters described in
Section~\ref{subsection::Simulation_Study} are used in separate
simulations with IMF clock angles of~$120^\circ$
(Fig.~\ref{fig::120_Dipole_Tilt}) and~$30^\circ$
(Fig.~\ref{fig::30_Dipole_Tilt}).  Figures~\ref{fig::120_Dipole_Tilt}
and~\ref{fig::30_Dipole_Tilt} display each model's prediction as the
grey squares, where the maximized quantity is: (a) the magnetic shear
angle $\theta$ in degrees, (b) the asymmetric outflow speed
$c_{A,out}$ in km/s, (c) the asymmetric Sweet-Parker reconnection rate
$E_{SP}$ in mV/m, (d) the quantity $B_\text{MS,L}^2B_\text{SH,L}^2$
from Eq.~\eqref{eqn::E_Hesse} in $10^8$~nT$^4$, and (e) the current
density magnitude $\left|\mathbf{J}\right|$ in nA/m$^2$, and (f) the
projection of the uniform $B_M$ component in the $x=0$ plane.  The
magnetic separators are displayed in white for (a)-(e) and black in
(f) as before.

In Fig.~\ref{fig::120_Dipole_Tilt}, each model appears to map the
magnetic separator reasonably well with a southward IMF orientation
and positive dipole tilt.  The maximum magnetic shear model's
prediction follows the separator more closely than any other model
when the dipole tilt is present.  The other models map the separator
more closely in the northern, dusk quadrant in
Figs.~\ref{fig::120_Dipole_Tilt}(b)-(e) than any other quadrant. [We
use dawn $\left(-y\right)$, dusk $\left(+y\right)$, southern
$\left(-z\right)$,, northern $\left(+z\right)$ to indicate position in
the $y$-$z$ GSM plane.]  Figure~\ref{fig::120_Dipole_Tilt}(f) shows
that the uniform $B_M$ component model again fails to map the magnetic
separator in any quadrant.

The results for IMF clock angle 30$^\circ$ and positive dipole tilt in
Fig.~\ref{fig::30_Dipole_Tilt} are significantly different.  The
separator moves approximately 45$^\circ$ in longitude duskward of the
$+x_\text{GSM}$ axis.  The model predictions intersect the magnetic
separator in the northern, dusk and southern, dawn quadrants, with the
maximum current density mapping the largest portion of the separator
in the northern, dusk quadrant.  The maximum magnetic shear model is
again the only model that faithfully reproduces the nightward portion
of the magnetic separator.  Most importantly, no model maps any part
of the separator in the southern, dusk quadrant.

\subsection{Effect of a Flow Shear}
\label{subsection::Shear_Flow}

Existing models have thus far ignored the impact of solar wind flow
around the magnetopause.  The tailward flow of magnetosheath plasma
around the magnetopause can have a component parallel to the
reconnecting component of the magnetic field, which is known to impact
the efficiency of
reconnection~\citep{chen1990,labelle-hamer1994,labelle-hamer1995,cassak2011b}.
This effect potentially could alter models based on reconnection
efficiency.

The incorporation of a plasma flow shear into
equations~\eqref{eqn::Asymmetric_Outflow_Speed}
and~\eqref{eqn::SP_Asymmetric_Reconnection_Rate} is ongoing work.
Based on preliminary work~(C.~E.~Doss et al., private communication,
2014), we hypothesize that the predictions for the outflow
speed~\citep{cassak2011a} and reconnection rate~\citep{cassak2011b}
for symmetric reconnection with a flow shear can be generalized for
asymmetric reconnection by replacing the symmetric outflow speed $c_A$
with the asymmetric outflow speed $c_{A,out}$ from
Eq.~\eqref{eqn::Asymmetric_Outflow_Speed}, giving
\begin{equation}
  c_\text{Shearflow}=c_{A,out}\sqrt{1-\frac{v_s^2}{c_{A,out}^2}},
\label{eqn::v_Shearflow}
\end{equation}
and
\begin{equation}
  E_\text{Shearflow}=E_{SP}\left(1-\frac{v_s^2}{c_{A,out}^2}\right),
\label{eqn::E_Shearflow}
\end{equation}
where $E_{SP}$ is given by
Eq.~\eqref{eqn::SP_Asymmetric_Reconnection_Rate}. The flow shear speed
$v_s$ is given by
\begin{equation*}
  v_s=\frac{v_\text{SH,L}-v_\text{MS,L}}{2}
\end{equation*}
and $v_\text{SH,L}$ and $v_\text{MS,L}$ are the magnetosheath and
magnetopause flow in the $L$ direction, respectively.

We test to see if these expression alter the predictions of the
reconnection efficiency arguments.  As before, $\alpha$ is discretized
to calculate the asymmetric arguments $c_{A,out}$ and $E_{SP}$ while
the $L$ direction is rotated around the magnetopause normal for each
$\alpha$.  The velocities on each side of the magnetopause are
measured 3$\delta$ upstream along $\mathbf{\hat{n}}$ and
$v_\text{SH,L}$ and $v_\text{MS,L}$ are calculated by taking the
$L$-component for each $\alpha$.  We calculate the modified
reconnection outflow speed and electric field using
equations~\eqref{eqn::v_Shearflow} and~\eqref{eqn::E_Shearflow} as a
function of $\alpha$. (For $v_s>c_{A,out}$, $c_\text{Shearflow}$ and
$E_\text{Shearflow}$ are set to zero as reconnection should not be
possible.)  Modified model predictions are determined with image
processing techniques as described in
Section~\ref{subsection::Ridge_Detection}.

\begin{figure*}
\centering
\noindent\includegraphics[width=39pc]{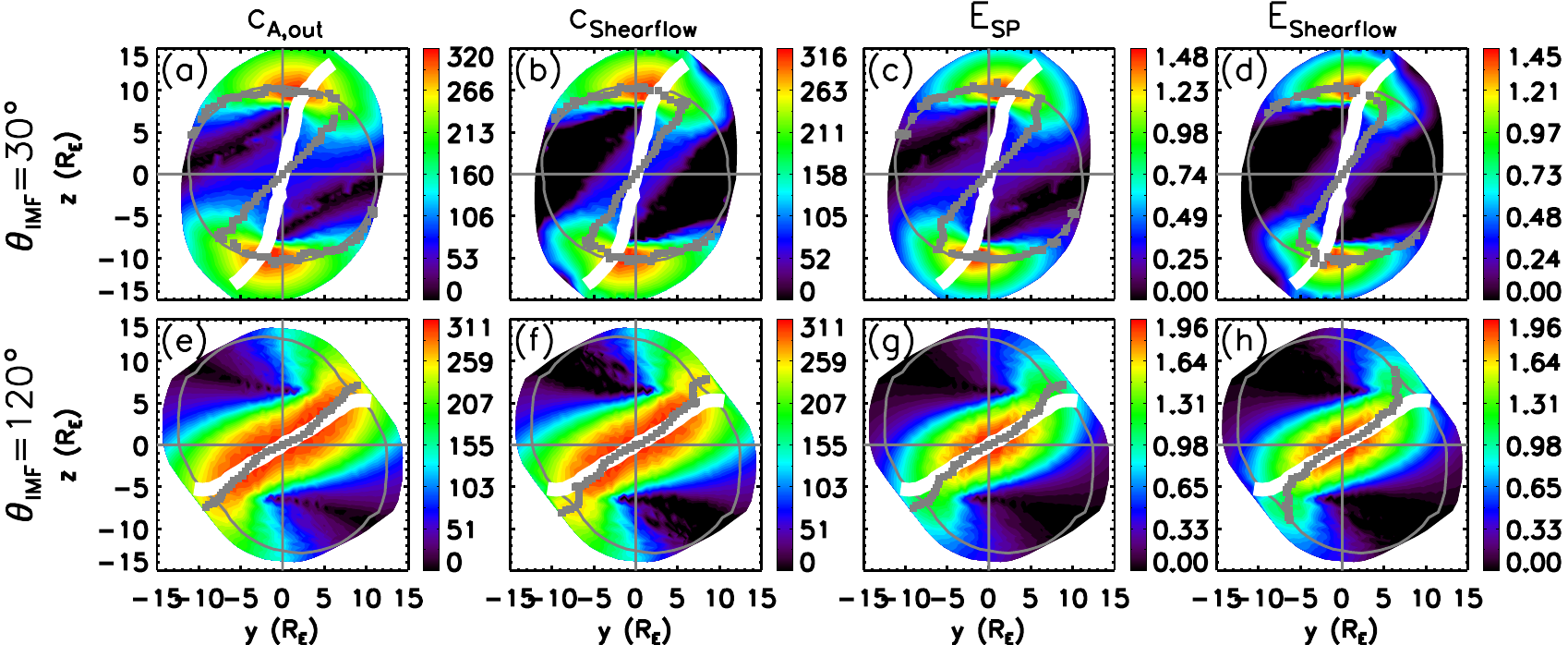}
\caption{Results for testing the effects of flow shear on model
  predictions with (top row) $\theta_\text{IMF}=30^\circ$ and (bottom
  row) 120$^\circ$.  Plots (a) and (e) are for the unmodified
  asymmetric outflow speed $c_{A,out}$, (b) and (f) are for the
  modified outflow speed $c_\text{Shearflow}$, (c) and (g) are for the
  unmodified asymmetric Sweet-Parker reconnection rate $E_{SP}$, and
  (d) and (h) are for the modified reconnection rate
  $E_\text{Shearflow}$. See the caption of
  Figure~\ref{fig::Magnetic_Shear} for definitions.}
\label{fig::Shearflow_Test}
\end{figure*}

Figure~\ref{fig::Shearflow_Test} displays results for two different
clock angles: the top row are from the simulation with
$\theta_\text{IMF}=30^\circ$ and the bottom row has
$\theta_\text{IMF}=120^\circ$.  Panels~(a) and~(e) are results without
corrections for flow shear and are replicated from
Fig.~\ref{fig::Outflow_Speed}. Likewise, panels~(c) and~(g) are from
Fig.~\ref{fig::Reconnection_Rate} for the reconnection rate. Panels
(b) and (f) give results that incorporate a flow shear in the
calculation of the outflow speed and (d) and (h) on the reconnection
rate. The model predictions more closely map the separator when the
IMF has a northward orientation after accounting for the flow shear,
and remain largely unchanged with a southward IMF orientation.
Quantifying this effect in the northward IMF simulation, the
prediction for $c_{A,out}$ makes an angle $\alpha$ of 44.7$^\circ$
with the magnetospheric magnetic field, and changes to 32.7$^\circ$
after accounting for the flow shear; the separator makes an angle of
$\alpha_{Sep}=10.3^\circ$.  Likewise, the prediction of $E_{SP}$ makes
an angle of 43.2$^\circ$ and is 29.7$^\circ$ after including a flow
shear.  Changes to the orientation angles for each model's prediction
in the southward IMF simulation are much smaller changing from
47.0$^\circ$ to 46.4$^\circ$ for the outflow speed, and change from
51.0$^\circ$ to 52.1$^\circ$ for the reconnection electric field; the
separator has $\alpha_{Sep}=55.3^\circ$ for this simulation.  A
further test of this model was performed for the 30$^\circ$ clock
angle with positive dipole tilt simulation.  The results (not shown)
revealed improvement of the model predictions in some areas, and worse
agreement in others, but still differed significantly from the
separator.

The relative importance of a flow shear is understood as a result of
general flow patterns around the magnetopause.  For southward IMF,
reconnection occurs near the ecliptic plane where the magnetosheath
flow and magnetospheric convection are tangential to this plane.
Thus, any flow shear parallel to the magnetic field is small and
reconnection should not be strongly affected.  However, the
magnetosheath flow at high latitudes is quite large.  When the IMF is
northward, there is appreciable flow parallel to the reconnecting
magnetic fields and may lead to the suppression of reconnection. The
present results suggest that including flow shear improves the
agreement between separators and model predictions for the
reconnection efficiency arguments.

We note that the analysis here assumes that any $v_M$ flow tangential
to the $L$-components of the magnetic fields has no effect on the
reconnection.  This is unlikely to be the case, but including this is
beyond the scope of the present study.

\section{Conclusions}
\label{section::Conclusions}
In this paper, we calculate the predictions of several magnetic
reconnection location models at the dayside magnetopause in global
resistive MHD simulations.  Models tested are: the maximum magnetic
shear model~\citep{trattner2007}, maximization of the asymmetric
reconnection outflow speed~\citep{swisdak2007}, maximization of the
asymmetric reconnection rate~(Shay, private communication,
2009;~\citep{borovsky2013}), the angle of
bisection~\citep{moore2002,borovsky2008b,hesse2013}, the maximization
of the current density magnitude~\citep{alexeev1998}, and component
reconnection (uniform $B_M$
component)~\citep{sonnerup1974,gonzalez1974}.  The first five models
predict reconnection to occur at locations where a given parameter is
maximized.  We employ robust image processing techniques to determine
each model's prediction.  Each model is tested in separate global
magnetospheric simulations with various IMF orientations, IMF
strengths, and with a dipole tilt.  The predictions for each model are
then compared to the magnetic separators to determine which model, if
any, accurately predicts the location of reconnection at the dayside
magnetopause.  We summarize our findings:
\begin{enumerate}
\item For the chosen simulation parameters, we find that all models
  are within a few Earth radii $\left(\text{R}_\text{E}\right)$ of the
  magnetic separators when the IMF has a southward orientation and no
  dipole tilt.
\item None of the models chosen for this study faithfully reproduce
  the entire magnetic separators when the IMF has a northward
  orientation and no dipole tilt.  However, only the maximum magnetic
  shear model faithfully reproduces the portion of the separator
  nightward of the magnetic nulls where reconnection is expected to occur.
\item The asymmetric outflow speed and asymmetric Sweet-Parker
  reconnection rate more closely map the magnetic separators than
  other models tested under southward IMF orientations and no dipole
  tilt.  The maximum magnetic shear model has a fixed orientation at
  the subsolar magnetopause and does not rotate with the magnetic
  separators for different IMF clock angles in simulations without a
  dipole tilt.
\item As the IMF strength is varied with fixed clock angle, the
  magnetic separators display different orientations than the
  predictions of the magnetic shear angle, and the asymmetric outflow
  speed and Sweet-Parker reconnection rate models.
\item In simulations with a dipole tilt of $+15^\circ$, we find most
  of the models map the magnetic separators when the IMF clock angle
  is 120$^\circ$.  However, none of the models chosen for this study
  faithfully reproduce the entire magnetic separator for a northward
  IMF orientation of 30$^\circ$.  Again, only the maximum magnetic
  shear model traces the portion of the separator nightward of the
  magnetic nulls.  The dayside portion of the magnetic separator in
  this simulation moves duskward by approximately 45$^\circ$ in
  longitude, as measured from the $+x_\text{GSM}$ axis.
\item Accounting for flow shear parallel to the reconnecting
  components of the magnetosheath and magnetospheric magnetic fields
  improve the performance of the asymmetric outflow speed and
  asymmetric reconnection rate models under northward IMF
  orientations.  Predictions for southward IMF orientations are mostly
  unaffected.
\end{enumerate}

In short, most of the models tested give results fairly close to the
location of dayside reconnection for southward IMF orientations, but
none are perfect.  We conclude that the existing models may be good
enough for a ball park estimate, but it remains unclear what
physically controls the location of magnetic reconnection at the
dayside magnetopause.

The present study employed a few underlying assumptions.  We employed
a uniform, explicit resistivity in our global MHD simulations even
though Earth's magnetopause is collisionless; this choice ensures our
simulations are well resolved, while reducing the likelihood of Flux
Transfer Events (FTEs)~\citep{russell1978} in our
simulations. However, recent advances have been made to trace magnetic
separators in simulation when FTEs are present~(A.~Glocer et al.,
manuscript in preparation, 2014).

Most of the reconnection location models discussed here are
independent of dissipation mechanism, so the present results on the
validity of each model should be independent of our choice to include
an explicit resistivity.  However, it is still unknown if the magnetic
separators depend on dissipation mechanism.  Future work will need to
extend the present study to global Hall-MHD, hybrid, or kinetic
simulations.

The analysis presented here has largely focused on the global process
of magnetic reconnection by identifying the magnetic separators across
the dayside magnetopause. At present, there is debate as to whether
magnetic reconnection is a global or local phenomenon at the dayside
magnetopause.  We determined a model's prediction by maximizing a
parameter related to reconnection at the dayside magnetopause.  Our
underlying assumption is that a particular model determines
reconnection if that model faithfully reproduces the entire magnetic
separator.  Future work will need to assess this assumption by
characterizing the locations on the magnetic separator where magnetic
reconnection occurs.

Finally, the present study uses a small set of solar wind and
magnetospheric conditions.  The present work has ignored the effect of
including an IMF $B_x$ component.  Previous studies found that under
southward IMF orientations, the reconnection site moves northward for
$B_x>0$ and southward when $B_x<0$~\citep{peng2010,hoilijoki2014}.
Additionally, we perform our analysis after the simulations have
achieved steady-state, which does not capture the magnetosphere's
response to dynamic solar wind
conditions~\citep{laitinen2006,laitinen2007}.  Understanding the
response of Earth's magnetosphere for a broader range of solar wind
conditions is of the utmost importance for realistic space weather
forecasting, and will be the subject of future work.

\begin{acknowledgments}

  Support from NSF grant AGS-0953463 (CMK and PAC), NASA grant
  NNX10AN08A (PAC), and NASA West Virginia Space Grant Consortium
  (CMK) are gratefully acknowledged.  We wish to acknowledge support
  from the International Space Science Institute in Bern, Switzerland.
  Simulations were performed at the Community Coordinated Modeling
  Center at Goddard Space Flight Center through their public Runs on
  Request system (http://ccmc.gsfc.nasa.gov). The CCMC is a
  multi-agency partnership between NASA, AFMC, AFOSR, AFRL, AFWA,
  NOAA, NSF and ONR. The BATS-R-US Model was developed by the Center
  for Space Environment Modeling at the University of Michigan. The
  analysis presented here was made possible via the Kameleon and Space
  Weather Explorer software packages provided by the Community
  Coordinated Modeling Center at NASA Goddard Space Flight Center
  (http://ccmc.gsfc.nasa.gov). Software Developers are: M.~M.~Maddox,
  D.~H.~Berrios, and L.~Rastaetter.  The data used to produce the
  results of this paper are publicly available for free from CCMC.
  The authors would like to thank C.~E.~Doss, S.~A.~Fuselier,
  G.~Paschmann, S.~M.~Petrinec, D.~G.~Sibeck, V.~M.~Souza, and
  K.~J.~Trattner for their insight and interesting discussions.

\end{acknowledgments}




\end{article}
\end{document}